\documentclass[journal]{IEEEtran}

\usepackage{amsmath,amsfonts}
\usepackage{algorithmic}
\usepackage{algorithm}
\usepackage{array}
\usepackage{textcomp}
\usepackage{stfloats}
\usepackage{url}
\usepackage{verbatim}
\usepackage{graphicx}
\usepackage{etoolbox}
\usepackage{cite}
\usepackage{threeparttable} 
\usepackage{multirow} 
\usepackage{xcolor}
\usepackage{enumerate}
\usepackage{amsthm}
\usepackage{array}
\usepackage{booktabs} 

\usepackage{subfigure}
\newcommand{\MyMapTemplatePrefix}[4]{\expandafter#1\csname#3#4\endcsname{#2{#4}}}
\newcommand{\MyMapTemplatePrefixNew}[5]{\expandafter#1\csname#4#5\endcsname{#2{#3{#5}}}}
\forcsvlist{\MyMapTemplatePrefix {\def} {\mathbf} {}} {A,B,C,D,E,F,G,H,I,J,K,L,M,N,O,P,Q,R,S,T,U,V,W,X,Y,Z}
\forcsvlist{\MyMapTemplatePrefix {\def} {\mathbf} {}} {a,b,c,d,e,f,g,h,i,j,k,l,m,n,o,p,q,r,s,t,u,v,w,x,y,z,1,0}
\forcsvlist{\MyMapTemplatePrefix {\def} {\widetilde} {wt}} {A,B,C,D,E,F,G,H,I,J,K,L,M,N,O,P,Q,R,S,T,U,V,W,X,Y,Z}
\forcsvlist{\MyMapTemplatePrefix {\def} {\widetilde} {wt}} {a,b,c,d,e,f,g,h,i,j,k,l,m,n,o,p,q,r,s,t,u,v,w,x,y,z} 
\forcsvlist{\MyMapTemplatePrefixNew {\def} {\widetilde}{\mathbf} {tb}} {A,B,C,D,E,F,G,H,I,J,K,L,M,N,O,P,Q,R,S,T,U,V,W,X,Y,Z}
\forcsvlist{\MyMapTemplatePrefixNew {\def} {\widetilde}{\mathbf} {tb}} {a,b,c,d,e,f,g,h,i,j,k,l,m,n,o,p,q,r,s,t,u,v,w,x,y,z}
\forcsvlist{\MyMapTemplatePrefix {\def} {\widehat} {wh}} {A,B,C,D,E,F,G,H,I,J,K,L,M,N,O,P,Q,R,S,T,U,V,W,X,Y,Z}
\forcsvlist{\MyMapTemplatePrefix {\def} {\widehat} {wh}} {a,b,c,d,e,f,g,h,i,j,k,l,m,n,o,p,q,r,s,t,u,v,w,x,y,z}
\forcsvlist{\MyMapTemplatePrefixNew {\def} {\widehat}{\mathbf} {hb}} {A,B,C,D,E,F,G,H,I,J,K,L,M,N,O,P,Q,R,S,T,U,V,W,X,Y,Z}
\forcsvlist{\MyMapTemplatePrefixNew {\def} {\widehat}{\mathbf} {hb}} {a,b,c,d,e,f,g,h,i,j,k,l,m,n,o,p,q,r,s,t,u,v,w,x,y,z}
\forcsvlist{\MyMapTemplatePrefixNew {\def} {\overline}{\mathbf} {lb}} {A,B,C,D,E,F,G,H,I,J,K,L,M,N,O,P,Q,R,S,T,U,V,W,X,Y,Z}
\forcsvlist{\MyMapTemplatePrefixNew {\def} {\overline}{\mathbf} {lb}} {a,b,c,d,e,f,g,h,i,j,k,l,m,n,o,p,q,r,s,t,u,v,w,x,y,z}
\forcsvlist{\MyMapTemplatePrefix {\def} {\mathcal}{mc}} {A,B,C,D,E,F,G,H,I,J,K,L,M,N,O,P,Q,R,S,T,U,V,W,X,Y,Z}
\forcsvlist{\MyMapTemplatePrefix {\def} {\mathbb} {mb}} {A,B,C,D,E,F,G,H,I,J,K,L,M,N,O,P,Q,R,S,T,U,V,W,X,Y,Z}

\def\utu{\mathrm{U2U}} \def\iti{\mathrm{I2I}}
\def\utt{\mathrm{user}} \def\itt{\mathrm{item}}
\usepackage{amsmath,amsfonts}
\usepackage{algorithmic}
\usepackage{algorithm}
\usepackage{array}
\usepackage[caption=false,font=normalsize,labelfont=sf,textfont=sf]{subfig}
\usepackage{textcomp}
\usepackage{stfloats}
\usepackage{url}
\usepackage{verbatim}
\usepackage{graphicx}
\usepackage{cite}
\begin{document}

\title{HeLLM: Multi-Modal Hypergraph Enhanced LLM Learning for Recommendation}

\author{Xu Guo, Tong Zhang, Yuanzhi Wang, Chenxu Wang, Fuyun Wang, 

Xudong Wang, Xiaoya Zhang, Xin Liu and Zhen Cui
\thanks{
Xu Guo, Tong Zhang, Yuanzhi Wang, Chenxu Wang, Fuyun Wang, Xudong Wang, and Xiaoya Zhang are with the School of Computer Science and Engineering, Nanjing University of Science and Technology, Nanjing, China. Email: \{xu.guo, tong.zhang, yuanzhiwang, 122106222830, fyw271828\}@njust.edu.cn}

\thanks{Xin Liu is with Shituoyun (Nanjing) Technology Co., Ltd, Nanjing, China.    Email: xin.liu@seetacloud.com.}

\thanks{Zhen Cui is with the School of Artificial Intelligence, Beijing Normal University, Beijing, China.  Email: zhen.cui@bnu.edu.cn.
}

\thanks{Corresponding author: Zhen Cui.}
}

\markboth{Journal of \LaTeX\ Class Files,~Vol.~14, No.~8, August~2021}%
{Shell \MakeLowercase{\textit{et al.}}: A Sample Article Using IEEEtran.cls for IEEE Journals}


\maketitle

\begin{abstract}
    The burgeoning presence of Large Language Models (LLM) is propelling the development of personalized recommender systems. 
    Most existing LLM-based methods fail to sufficiently explore the multi-view graph structure correlations inherent in recommendation scenarios. 
    To this end, we propose a novel framework, \underline{H}ypergraph \underline{E}nhanced \underline{LLM} Learning for multimodal Recommendation (HeLLM), designed to equip LLMs with the capability to capture intricate higher-order semantic correlations by fusing graph-level contextual signals with sequence-level behavioral patterns. 
    In the recommender pre-training phase, we design a user hypergraph to uncover shared interest preferences among users and an item hypergraph to capture correlations within multimodal similarities among items.  
    The hypergraph convolution and synergistic contrastive learning mechanism are introduced to enhance the distinguishability of learned representations. 
     During the LLM fine-tuning phase, we directly inject the learned graph-structured embeddings into the LLM’s architecture and integrate sequential features that capture each user’s chronological behavior. 
    This process enables hypergraphs to leverage graph-structured information as global context, enhancing the LLM’s ability to perceive complex relational patterns and integrate multimodal information, while also modeling local temporal dynamics.
    Extensive experiments demonstrate the superiority of our proposed method over state-of-the-art baselines, confirming the advantages of fusing hypergraph-based context with sequential user behavior in LLMs for recommendation. 
    The code is publicly available at \url{https://github.com/Xu107/HeLLM}. 
\end{abstract}


\section{Introduction}
\IEEEPARstart{R}{ecommender} systems aim to mitigate information overload by uncovering and satisfying users' implicit preferences and needs ~\cite{ying2018graph,he2020lightgcn}. 
Large Language Models (LLM) have demonstrated unprecedented capabilities across various domains like NLP \cite{achiam2023gpt}, Computer Vision ~\cite{wang2024visionllm,li2023blip}, and Medicine \cite{thirunavukarasu2023large}, and recent efforts have explored their potential to enhance the personalization of recommendation. 

\begin{figure}[!tbp]
	\includegraphics[width=1.0\linewidth]{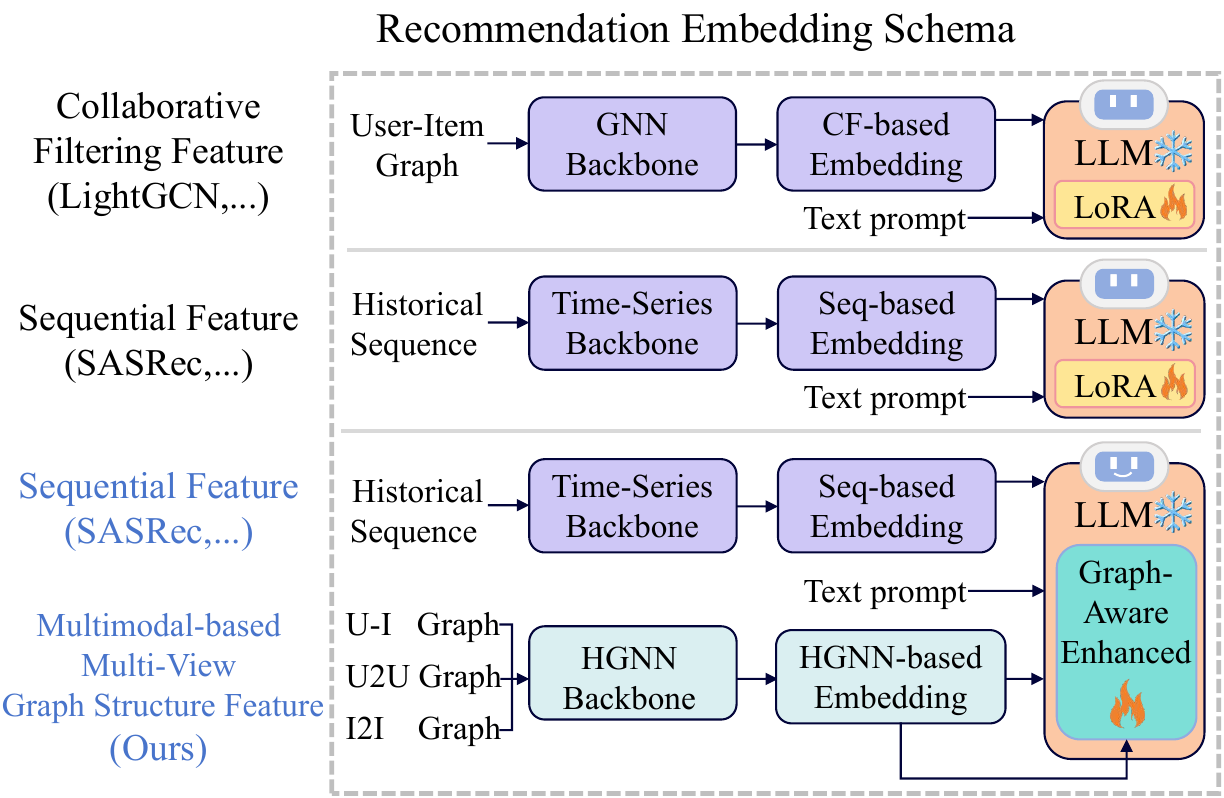}
	\caption{
    Comparison of existing LLM-based models and our proposed method, which combines the strengths of GNN and time-series approaches to capture richer multimodal, multi-view graph structures, enhancing the LLM's ability to perceive and reason over complex relational patterns. 
	}
	\label{fig:main_idea}
\end{figure}

Recent studies ~\cite{guo2023automated,wu2024exploring,li2023personalized,zhang2023collm,liao2024llara,lin2024data,wu2024personalized} employ Parameter-Efficient Fine-Tuning (PEFT) ~\cite{ding2023parameter,lialin2023scaling} to bridge the knowledge gap between LLM and recommender models by optimizing selective or additional lightweight parameters. 
These models are typically structured within a Recommendation Embedding Schema, consisting of a pre-trained lightweight recommendation backbone that learns embeddings for users and items, a linear mapping layer that aligns these embeddings with the LLM’s representation space, and an LLM backend with textual prompt for enriching semantic understanding. 

As shown in Fig.~\ref{fig:main_idea}, existing methods fall into two main categories: the first ~\cite{zhang2023collm} represents whole user-item interactions as a bipartite graph, utilizing Graph Neural Network (GNN) backbones to capture Collaborative Filtering (CF) signals; the second ~\cite{li2023e4srec,liao2024llara} models each user's interaction history with a time-series backbone to capture user's sequential behavior patterns. 
For example, Collm \cite{zhang2023collm} integrates learned collaborative representations of LightGCN \cite{he2020lightgcn} into LLM for recommendation. 
LLaRA \cite{liao2024llara} harness traditional sequential recommenders such as GRU4Rec \cite{hidasi2015session} and SASRec \cite{kang2018self} to encode “sequential behaviors of users" as a distinct modality beyond texts. 
These methods have shown promising results in integrating user interaction patterns with LLM-driven semantics understanding. 

However, in practical scenarios, both GNN-based and time-series approaches are essential, each capturing unique facets of user behavior that are indispensable for the comprehensive recommendation. 
GNN-based CF methods ~\cite{wang2019neural,he2020lightgcn,wu2022graph} explicitly model intricate relational structures across the entire user-item interaction space, capturing multi-hop dependencies among users and items via message propagation, and providing a global perspective on interactions. 
But they are limited in capturing the fine-grained temporal evolution, which is crucial for modeling users' time-sensitive sequential behaviors. 
Conversely, sequence-based methods \cite{kang2018self,zhou2020s3,wu2024personalized} capture temporal dependencies yet fail to adequately mine abundant structured correlations (e.g., social networks) among users and items within the recommendation context, lacking a global CF supervisory signal.

Evidently, LLM-based methods hold the potential to improve performance by increasing the power of pre-trained representations. 
Despite the complementary advantages of CF-based and sequence-based embeddings, their mutual effect remains largely unexplored. 
Existing LLM-based methods ~\cite{liao2024llara,zhang2023collm} primarily focus on learnable ID attributes and user-item bipartite graphs, also with limited exploration of high-order shared user preferences and multimodal side information correlations among items. 
Typically, these methods leverage enhanced prompt learning \cite{lester2021power} (tied to pre-trained embeddings) and additional LoRA ~\cite{hu2021lora} parameters (independent of pre-trained embeddings) for PEFT of LLM. Recent study \cite{ding2023parameter} indicates that fine-tuning intermediate layers (e.g., via LoRA) yields superior performance compared to adjustments in the initial (prompt learning-like) or final layers. 
This insight motivates us to inject rich graph-structured information into LLM internals during fine-tuning, aiming to unlock graph-perceived potential. 
In summary, we identify two core objectives for advancing LLM-based recommender systems: (1) enhancing the expressive power of pre-trained representations, and (2) effectively integrating rich graph structure information to strengthen the LLM’s ability to perceive relational patterns.

    To this end, we shift our attention to enhancing graph-structured representation and improving LLM's ability to perceive and reason over complex graph structures. 
    We propose a novel framework, \underline{H}ypergraph \underline{E}nhanced \underline{LLM} Learning for Multimodal Recommendation (HeLLM), designed to comprehensively capture multi-view graph structures, enhancing LLM for more effective user recommendation. 
    Accordingly, a two-phase learning process, including recommender pre-training and LLM fine-tuning, is designed to mine underlying user preferences and inject them into LLM progressively.
    In the pre-training phase, we attempt to adequately model the abundant graph structures in the multi-modal recommendation and to learn discriminative graph-based representation. 
    Specifically, we construct two hypergraphs: a user-to-user (U2U) hypergraph to extract shared user preferences and an item-to-item (I2I) hypergraph to capture intricate semantic multi-modal resemblance among items, as complementary to the primary first-order user-item interaction. 
    By doing so, we can transform multimodal side information of items into a hypergraph structure. 
    Then, hypergraph convolution is conducted to extract second-order semantic information on the user and item sides separately. 
    Meanwhile, the synergistic contrastive learning mechanism is designed to maximize the mutual information between first-order and second-order representations of users/items, improving the distinguishability of features.  
    During LLM fine-tuning, we introduce a graph-aware enhancement mechanism that incorporates multi-view graph-based features. First, pooling operations on graph embeddings capture global structural patterns for direct injection into the LLM's interior. Then, time-series and graph embeddings are fused via a multilayer perceptron to align the recommender and LLM feature spaces. 
    In this way, LLM can effectively excavate intricate higher-order semantic correlations among users and items, and utilize graph-based features as a bridge to integrate multi-modal information. 
    The extensive experiments confirm the superiority of our proposed method over state-of-the-art baselines. 
    Our key contributions can be summarized as follows:
    \begin{itemize}
        \item We propose the HeLLM, a novel framework that enhanced LLM with time-series and multi-view graph structural embeddings to capture user temporal behaviors and high-order user-item correlations for recommendation. 
        \item We introduce an effective pretraining recommender component that leverages hypergraph convolution and a synergistic contrastive learning mechanism to explore intricate higher-order shared preferences among users (via a U2U hypergraph) and potential multimodal semantic correlations among items (via an I2I hypergraph). 

        \item We design an internal graph-aware enhancement mechanism that improves LLM with the ability to perceive and reason over complex graph structures. 
        \item Extensive experiments on three public datasets validate the superiority of our proposed method over various state-of-the-art multimodal recommendation baselines. 
    \end{itemize}

\section{Related Work}
\label{sec:formatting}
\begin{figure*}[!htp]
    \centering
    \includegraphics[width=1\textwidth]{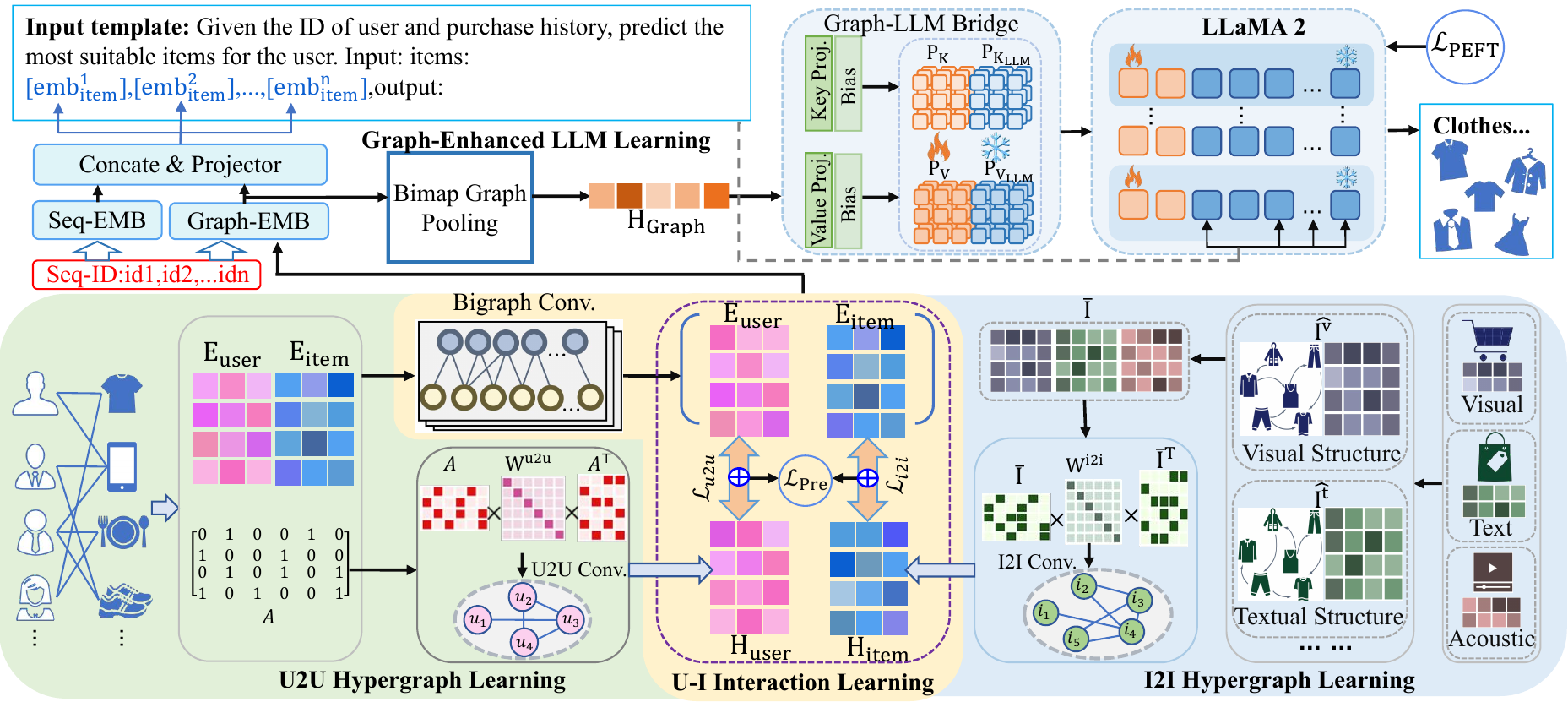}
    \caption{
    The architecture of our proposed HeLLM. The recommender pre-training phase (lower half of the figure) comprises three components: i) U2U Hypergraph Learning (green) for user-level representations; ii) I2I Hypergraph Learning (blue) leveraging multimodal item's side information for learning item-level representations; and iii) U-I Interaction Learning (yellow), focusing on user-item bigraph representations with integrated collaborative contrastive learning. In the LLM fine-tuning phase (upper half of the figure), iv) Graph-Enhanced LLM Learning injects graph-based features into the LLM's internal structure and incorporates sequence embeddings as item representations. For further details, please see the corresponding sections. 
    }
    \label{fig:framework}
\end{figure*}
\subsection{GNN-based Recommender Systems.} 
Graph neural networks (GNN) are widely used in recommender systems to model graph-structured data and learn representations~\cite{wu2022graph}. 
    These methods generally fall into five categories: 
    1) Collaborative filtering based on user-item interactions (e.g., NGCF~\cite{wang2019neural}, LightGCN~\cite{he2020lightgcn}). 
    2) Sequential and session recommendation based on time-series user historical behaviors (e.g., SR-GNN~\cite{wu2019session}, GC-SAN~\cite{xu2019graph}). 
    3) Social recommendation based on user associations (e.g.,  ESRF~\cite{yu2020enhancing}, GraphRec ~\cite{fan2019graph}).  
    4) Knowledge graph based recommendation ~\cite{wang2019knowledge,wang2019kgat}.
    5) Multimodal based recommendation~\cite{he2016vbpr,wei2019mmgcn,wei2020graph,zhang2021mining,wang2021dualgnn,yi2021multi,zhou2023tale,chen2017attentive,chen2019personalized,wei2023multi,zhou2023bootstrap,tao2022self,guo2025mmhcl,guo2025m}. 
    For example, VBPR \cite{he2016vbpr} enhances matrix factorization by incorporating visual features alongside item ID representations. 
    MMGCN \cite{wei2019mmgcn} initiated the use of graph-based modeling in multimodal recommendation by applying graph convolution networks to each modality.
    GRCN \cite{wei2020graph} mitigates the impact of false-positive signals by detecting and pruning noisy connections within the interaction graph. 
    In LATTICE \cite{zhang2021mining}, deep networks are utilized to extract implicit item structure. FREEDOM \cite{zhou2023tale} fixes these learned item structures and denoises the user-item interaction graph. 
    AlignRec \cite{liu2024alignrec} mitigates semantic misalignment via three-stage alignment among content, ID, and user-item features. 
    PGL \cite{yu2025mind} constructs principal subgraphs via global- and local-aware operators to jointly exploit co-occurrence patterns and personalized signals. 
    In contrast to prior approaches, our method explores second-order semantic correlations at the user and item levels to enhance multimodal recommendation and enrich LLM capabilities to perceive and reason over complex graph structures.

\subsection{Parameter-Efficient Fine-Tuning.} Parameter-Efficient Fine-Tuning methods ~\cite{ding2023parameter,lialin2023scaling} only optimize a small subset of parameters or introduce additional ones in the LLM while keeping the majority frozen to cut down computational and storage costs compared with full-tuning. 
    These methods fall into four categories:
    1) Addtion-based methods, which introduce extra learnable parameters into LLM and only train them (e.g. Adapter \cite{houlsby2019parameter}, Prompt-Tuning \cite{lester2021power}, Prefix-Tuning \cite{li2021prefix}, Ladder-Side Tuning \cite{sung2022lst}). 
    2) Selection-based methods, which selectively fine-tune a limited subset of parameters in LLM (e.g. BitFit \cite{zaken2021bitfit}, DiffPruning \cite{guo2020parameter}, FishMask \cite{sung2021training}). 
    3) Reparametrization-based methods, which reparameterize parameters of LLM into a low-dimensional intrinsic subspace using low-rank transformation (e.g. LoRa \cite{hu2021lora}, Intrinsic SAID \cite{aghajanyan2020intrinsic}). 
    4) Hybrid-based methods, which incorporate various approaches (e.g. MAM Adapter \cite{he2021towards}, UniPELT \cite{mao2021unipelt}).

\subsection{Large Language Models in Recommendation.} 
    Large Language Models ~\cite{vaswani2017attention,fan2023recommender} use transformer-based deep learning architecture, combined with extensive corpus data and parameters for training, showcasing unparalleled capabilities in understanding and generating human-like language~\cite{achiam2023gpt,zhao2023survey}. 
    Recent LLM-based recommendation methods fall into two categories: 
    1) Non-tuning-based methods (In-context learning methods~\cite{hou2024large}) ~\cite{gao2023chat,liu2023chatgpt,zhang2023chatgpt,dai2023uncovering,hou2024large,wang2023zero,ren2023representation,he2023large,sanner2023large} which do not fine-tune the LLM and leverage the rich knowledge of the LLM for zero-shot and few-shot recommendations. 
    2) PEFT-based methods~\cite{bao2023tallrec,li2023prompt,cui2022m6,wu2024personalized,wang2022towards,guo2023automated,wu2024exploring,li2023personalized,zhang2024notellm} that fine-tune a small set of LMM weights or add a few extra trainable parameters to synchronize the knowledge of LLM and recommendation models. 
    For example, TALLRec \cite{bao2023tallrec}  fine-tunes the additional injected Low-Rank Adaptation (LoRa) \cite{hu2021lora}, which employs a simple low-rank matrix decomposition to parameterize the weights in LLM. 
    POD \cite{li2023prompt} applies prompt tuning by introducing a trainable prompt token to bridge between IDs and words. 
    LLaRA \cite{liao2024llara} integrates sequential recommendation representation with LLMs by treating user behaviors as a distinct modality, aligning ID-based embeddings with textual features through hybrid prompting to enhance behavioral reasoning in LLMs. 
    CoLLM \cite{zhang2023collm} enhances LLM-based recommendation by integrating collaborative information from traditional models into LLMs' input space via collaborative embeddings. 
    Distinct from prior work, we explore multimodal-based multi-view graph structures to enhance the ability of representation, injecting them into LLM internals during fine-tuning to strengthen the LLM's capacity for perceiving complex relational patterns.

\section{The Proposed Method}

\subsection{Overview}

Given a target user and the historical interactions, the recommendation aim is to predict which item the user will interact with in the future, utilizing both the user’s past behavior and the items’ multimodal content. 
Let $\mathcal{U}$ and $\mathcal{I}$ represent the sets of $M$ users and $N$ items. The matrix
$\mathbf{A}\in\mbR^{M\times N}$ indicates historical interaction relationships between all users and items, where $\mathbf{A}_{u,i}=1$ if there is an interactive record $({u},{i})$\footnote{For simplification, we mix the use of $u$ and $i$, as one user/item or matrix/vector indices, in clear context.} for $ {u} \in \mathcal{U}$ and ${i} \in \mathcal{I}$, otherwise 0. 
In multimodal scenarios, items encompass diverse multimodal attributes. Here we use 
$\mathcal{M}=\{\mathbf{v},\mathbf{t}\}$ to denote a multimodal set comprising visual and textual modalities respectively. 


The whole architecture of our proposed HeLLM is illustrated in Fig.~\ref{fig:framework}, including the dual pipelines of the pre-training phase for learning user/item embeddings at the base and the fine-tuning phase for refining  LLM at the top. 
In the embedding learning phase, the inputs consist of user-item interactions displayed on the left, highlighted with a green background, and raw multimodal item features showcased on the right, distinguished by a blue background. 
This whole learning process integrates four pivotal learning modules: \textbf{i) User-to-User (U2U) Hypergraph Learning } constructs a U2U hypergraph from the entirety of user-item interactions to extract higher-order shared preferences among users and learn user-level representation. 
\textbf{ii) Item-to-Item (I2I) Hypergraph Learning} initially creates modal-specific graph structures based on raw feature similarity among items, which are then integrated into I2I hypergraph to unearth and refine intricate multimodal semantic associations, thus boosting item-level representation learning. 
\textbf{iii) U-I Interaction Learning} combines user-item representation from a bipartite graph with those from individual user and item sides, improving representation discrimination through collaborative contrastive learning. These fused representations serve as the embedding representation. 
In the LLM fine-tuning phase, \textbf{iv) Graph-Enhanced LLM Learning} pools global graph structural features for direct injection into the LLM’s interior, then concatenates time-series and graph embeddings through a multilayer perceptron to align with the LLM's feature space. 

\subsection{User-to-User (U2U) Hypergraph Learning} 

At the user level, we construct a U2U hypergraph, representing users as nodes and their item interactions as hyperedges to capture shared user preferences. This enables the exploration of second-order semantic relationships among users through hypergraph convolution, with items acting as bridges. The hypergraph encoder is constructed using stacked hypergraph convolutional layers, as in HGNN~\cite{feng2019hypergraph}, to learn user-level embeddings. Formally, 

\begin{equation}
    \mathbf{x}_{i}^{(l+1)}=\frac{1}{N} \sum_{j=1}^{M} \sum_{k=1}^{N} \mathbf{A}_{i k} \mathbf{A}_{j k} \mathbf{W}_{k k}^{u2u} \mathbf{x}_{j}^{(l)}, 
\end{equation} 

where $\mathbf{x}_{i}$ is the embedding of user $i$ and $l$ denotes the layer. The hyperedge weights $\mathbf{W}^{u2u}$ are trainable and can be initialized with a user-to-user similarity matrix, here we set it as an identity matrix for simplicity. The hypergraph convolution can be expressed in matrix form as:
\begin{equation}
    \begin{split}
         & \overline{\mathbf{H}}_{\utu} = \mathbf{LN} \left( \mathbf{A} \mathbf{W}^{u2u} \mathbf{A}^{\mathrm{T}} \right), \\
         & \mathbf{X}_{\utt}^{(l+1)} = \overline{\mathbf{H}}_{\utu} \mathbf{X}_{\utt}^{(l)}, \label{eqn:H_U2U}
    \end{split}
\end{equation}

where $\overline{\mathbf{H}}_{\utu}$ denotes the reorganized U2U hypergraph after convolution, 
$\mathbf{X}_{\utt}\in \mathbb{R}^{M \times d}$ denotes user embeddings initialized with the Xavier~\cite{glorot2010understanding}. 
Here, $\mathbf{LN}$ applies layer normalization (e.g., laplacian normalization) to prevent gradient explosion. The hypergraph structure facilitates a two-step information transfer: initially from nodes to hyperedges, and subsequently from hyperedges back to nodes. 
Specifically, in Eq.~(\ref{eqn:H_U2U}), $\mathbf{A}^{\mathrm{T}} \mathbf{X}_{\utt}^{(l)}$ gathers user information into item hyperedges, while the matrix product with $\mathbf{A}$ integrates this hyperedge information back into user nodes. 
The operation $\mathbf{A}\mathbf{A}^{\mathrm{T}}$ quantifies the shared item connections between users.
The final user-level embedding for a user $u$, denoted as $\mathbf{h}_{u}=[\X^{(L)}_{\utt}]_u$, is derived after $L$ iterations of hypergraph convolution.

\subsection{Item-to-Item (I2I) Hypergraph Learning} 

To capture complex higher-order semantic relationships and enhance multimodal information fusion among items, we construct an item-to-item (I2I) hypergraph at the item level. First, we compute cosine similarity within each modality to identify related item neighbors:
\begin{equation}
        \mathbf{I}_{ij}^m=({\mathbf{e}}_i^m)^\top{\mathbf{e}}_j^m/(\|{\mathbf{e}}_i^m\|_2\|{\mathbf{e}}_j^m\|_2),
\end{equation}
where $\mathbf{e}_{i}^{m} \in \mathbb{R}^{d_m}$ denotes the raw multimodal feature of item $i$ for modality $m \in \mathcal{M}$. The resulting correlation matrix $\mathbf{I}^m \in \mathbb{R}^{N \times N}$ captures intra-modality item relationships. To reduce redundant connections, we apply a K-nearest neighbors (KNN) filter, producing a sparse matrix:

\begin{equation}
        \widehat{\mathbf{I}}_{i j}^{m}=\left\{\begin{array}{ll}
        1, & \mathbf{I}_{i j}^{m} \in \text{ Top-$\mathrm{K}$}\left(\mathbf{I}_{i}^{m}\right), \\
        0, & \text{otherwise.}
        \end{array}\right.
\end{equation}
Then, we fuse multimodal information into the I2I hypergraph, denoted as $\overline{\mathbf{I}}={||}_{m}^{M} \widehat{\mathbf{I}}^m$, where $||$ represents concatenation across modalities. In the I2I hypergraph, items enriched with multimodal data serve as nodes, while the K-nearest neighbors in each modality form hyperedges. The I2I hypergraph convolution operates similarly to the U2U hypergraph, defined as:
    \begin{equation}
        \begin{split}
        & \overline{\mathbf{H}}_{\iti} = \mathbf{LN} \left( \overline{\mathbf{I}} \mathbf{W}^{i2i} \overline{\mathbf{I}}^{\mathrm{T}} \right), \\
        & \mathbf{X}_{\itt}^{(l+1) } = \overline{\mathbf{H}}_{\iti} \mathbf{X}_{\itt}^{(l)}, \label{eqn:H_I2I}
        \end{split}
    \end{equation}

    where $\overline{\mathbf{H}}_{\iti}$ represents the restructured I2I hypergraph post-convolution, $\mathbf{W}^{i2i}$ is similar to ${\mathbf{W}}^{u2u}$, and $\mathbf{X}_{\itt}$ are the item-level embeddings. This hypergraph structure enables capturing intricate higher-order item relationships. 
    After applying Eq.~(\ref{eqn:H_I2I}), the operation $\overline{\mathbf{I}}^{\mathrm{T}}\mathbf{X}_{\itt}^{(l)}$ consolidates modality-specific information from items nodes into hyperedges, enhancing intra-modal interactions. The subsequent multiplication by $\overline{\mathbf{I}}$ redistributes this hyperedge information back to items, facilitating inter-modal fusion. The output $\mathbf{h}_{i}=[\X^{(L)}_{\itt}]_i$ for one item $i$, obtained after using $L$ hypergraph convolutional layers, represents the final item-level embeddings.

    
\subsection{U-I Interaction Learning}

The integration of user-item representations derived from a bipartite interaction graph and those from individual user and item levels serves as the output for the recommender pre-training phase. We further design a collaborative contrastive learning mechanism to enhance the discriminative power of these representations. For efficiency, LightGCN \cite{he2020lightgcn} is employed to perform convolutions on the user-item bipartite graph. Let $\mathbf{e}_{u}$ and $\mathbf{e}_{i}$ denote the learned ID representations of user $u$ and item $i$, respectively. The normalized user-level $\mathbf{h}_{u}$ and item-level $\mathbf{h}_{i}$ representations are fused with $\mathbf{e}_{u}$ and $\mathbf{e}_{i}$ to enrich their information: 

\begin{equation}
        \widetilde{\mathbf{e}}_u =\mathbf{e}_u+\frac{\mathbf{h}_u}{\| \mathbf{h}_u\|_2}, 
        \widetilde{\mathbf{e}}_i =\mathbf{e}_i+\frac{\mathbf{h}_i}{\| \mathbf{h}_i\|_2},     
\end{equation}
where $\widetilde{\mathbf{e}}_u$ and $\widetilde{\mathbf{e}}_i$ represent the user and item learned from the training phase. The interaction probability between user $u$ and item $i$ is estimated with the calculation of inner product: $\hat{y}_{ui}=\langle\widetilde{\mathbf{e}}_u, \widetilde{\mathbf{e}}_i\rangle$. 
To improve representation distinguishability, we design collaborative contrastive learning between embeddings from macro-perspectives (user and item levels) and those from the collaborative filtering perspective. The loss functions for contrastive learning at both ends are defined as:  
    \begin{equation}
    \resizebox{.9\hsize}{!}{$
    \begin{aligned}
        & \mathcal{L}_{\mathrm{\utu}} =-\sum_{u\in\mathcal{U}}\log\frac{\exp (s(\mathbf{h}_u,\widetilde{\mathbf{e}}_u))}{\sum_{u^{\prime}\in\mathcal{U}}\left(\exp (s(\mathbf{h}_{u^{\prime}},\widetilde{\mathbf{e}}_u))+\exp (s(\widetilde{\mathbf{e}}_{u^{\prime}},\widetilde{\mathbf{e}}_u))\right)}, \\
        & \mathcal{L}_{\mathrm{\iti}} =-\sum_{i\in\mathcal{I}}\log\frac{\exp (s(\mathbf{h}_i,\widetilde{\mathbf{e}}_i))}{\sum_{i^{\prime}\in\mathcal{I}}\left(\exp( s(\mathbf{h}_{i^{\prime}},\widetilde{\mathbf{e}}_i))+\exp (s(\widetilde{\mathbf{e}}_{i^{\prime}},\widetilde{\mathbf{e}}_i))\right)}, 
        \end{aligned}
        $}
    \end{equation}
    where $s(\mathbf{x}_1,\mathbf{x}_2)=\mathbf{x}_1^\top\cdot\mathbf{x}_2/(\tau\cdot\|\mathbf{x}_1\|_2\cdot\|\mathbf{x}_2\|_2)$ is the similarity measurement with a temperature coefficient $\tau$. In these losses, the mutual information between hypergraph convolutional embeddings (e.g., shared preference patterns for users) and bigraph embeddings (information of selected items for users)  is optimized to maximize/minimize similarity for the same/different users and items. 

    We denote the final item embeddings, $\widetilde{\mathbf{E}}_{item}$, as the output of our recommender pre-training, representing graph-based embeddings ($\mathbf{E}^{Graph}$) for input to LLM fine-tuning.

\subsection{Graph-Enhanced LLM Learning}

    During LLM fine-tuning, the pre-trained graph-based embeddings $\mathbf{E}^{Graph}$ are integrated to enhance the LLM’s capacity for graph-structured reasoning. First, a global graph pooling operation extracts structural features for direct injection into the LLM’s interior: 
    \begin{equation}
        \overline{\mathbf{H}}_{\mathrm{G}}=\textbf{CONCAT}\left(\mathbf{E}_{(u)}^{Graph}\right).
    \end{equation}
    Here, $\mathbf{E}_{(u)}^{Graph}$ represents the collection of all items' graph features derived from the current interactions of the user ${u}$. 
    To extract graph-structured collaborative filtering signals and intricate higher-order semantic correlations from learned representations, a bilinear mapping second-order pooling similarly used in~\cite{wang2020second} is employed: 
    

    \begin{equation}
    \begin{split}
        & \widehat{\mathbf{H}}_{\mathrm{G}}=\mathbf{W}^{\mathrm{T}}{\overline{\mathbf{H}}}_{\mathrm{G}}^{\mathrm{T}}\overline{\mathbf{H}}_{\mathrm{G}}\mathbf{W}\in \mathbb{R}^{d^{\prime} \times d^{\prime}}; \\
        & \mathbf{H}_{\mathrm{Graph}}=\textbf{FLATTEN}\left( \widehat{\mathbf{H}}_{\mathrm{G}}\right) \in \mathbb{R}^{d^{\prime 2}},
    \end{split}
    \end{equation}
    where $d^{\prime}<d$ and $W \in \mathbb{R}^{f \times f^{\prime}}$ denotes a trainable matrix that performs a linear mapping, and the matrix is finally flattened to a feature vector as the global graph-structure representation. 
    Then, the graph structure vector is mapped into the feature space of LLM by using normalization and MLP transformation: 
    \begin{equation}
    \begin{split}
        & \widehat{\mathbf{P}}_{\mathrm{K}}=\textbf{MLP}_{\mathrm{K}}\left(\textbf{RMSNorm} \left(\mathbf{H}_{\mathrm{Graph}}\right)\right) \in \mathbb{R}^{\whd}, \\
        & \widehat{\mathbf{P}}_{\mathrm{V}}=\textbf{MLP}_{\mathrm{V}}\left(\textbf{RMSNorm} \left(\mathbf{H}_{\mathrm{Graph}}\right)\right) \in \mathbb{R}^{\whd},
        \end{split}
    \end{equation}
    where $\whd$ is the dimension of the large language model features.  
    $\widehat{\mathbf{P}}_{\mathrm{K}}$ and $\widehat{\mathbf{P}}_{\mathrm{V}}$ are graph-enhanced key and value representations that employ $k$ repetitions and add with learnable biases with broadcast to form $\mathbf{P}_{\mathrm{K}}$ and $\mathbf{P}_{\mathrm{V}}\in \mathbb{R}^{k \times \whd}$, respectively. 
    Subsequently, they integrate with frozen keys $\P_{\mathrm{K}_\mathrm{LLM}}$ and values $\P_{\mathrm{V}_\mathrm{LLM}}$ across each transformer layer of the LLM to conduct graph-structure-enhanced prefix-tuning:  
    \begin{equation}
    \begin{split}
        \P_{\mathrm{K}_\mathrm{LLM}}^{\prime}=\textbf{CONCAT}\left( \mathbf{P}_{\mathrm{K}},\P_{\mathrm{K}_\mathrm{LLM}}\right);
        \\
        \P_{\mathrm{V}_\mathrm{LLM}}^{\prime}=\textbf{CONCAT}\left( \mathbf{P}_{\mathrm{V}},\P_{\mathrm{V}_\mathrm{LLM}}\right).
        \end{split}
    \end{equation}
    In this way, the collaborative filtering signals as well as multimodal information are injected into the large language model through graph-based features as the bridge, thereby boosting the LLM's ability in graph-structured reasoning for recommendation tasks.
    
    Finally, each historical user record is converted into natural language, with item IDs replaced by concatenating with pre-trained graph-based embeddings and sequential-based embeddings from SASRec \cite{kang2018self}, the same as LLaRA \cite{   liao2024llara}. These embeddings are then aligned with the LLM’s feature space via a multilayer perceptron. In this work, we concentrate on processing graph-based embeddings and thus omit further details on time-series embeddings. 
     
\subsection{Objective Optimization}
The optimization objectives involve two stages: (1) extracting user and item representations from graph structures and multimodal data in the recommender pre-training phase, optimized with $\mathcal{L}_{\mathrm{Rec\_Pretrain}}$, and (2) fine-tuning the LLM by integrating graph-based and sequential features, guided by $\mathcal{L}_{\mathrm{LLM\_PEFT}}$. 
    
    Firstly, the loss function of learning recommender embeddings can be defined as:    
\begin{equation}
\mathcal{L}_{\mathrm{Rec\_Pretrain}}=\mathcal{L}_{\mathrm{BPR}}+\alpha\cdot\mathcal{L}_{\utu}+\beta\cdot\mathcal{L}_{\iti}+\lambda \cdot\|\Theta\|^{2},
\end{equation}
    where $\alpha$ and $\beta$ are the balance factors for $\mathcal{L}_{\mathrm{U2U}}$ and $\mathcal{L}_{\mathrm{I2I}}$, respectively for the two hypergraphs. The last term is the $L_{2}$ regularization with coefficient $\lambda$ to prevent overfitting. $\mathcal{L}_{\mathrm{BPR}}$ represents the Bayesian Personalized Ranking (BPR) loss \cite{rendle2012bpr}, which employs a pairwise approach to optimize the prediction of higher scores for observed items compared to unobserved ones, defined as: 
    \begin{equation}
        \mathcal{L}_{\mathrm{BPR}}=-\sum_{u=1}\sum_{i\in\mathcal{N}_{u}}\sum_{j\notin\mathcal{N}_{u}}\ln\sigma(\hat{y}_{ui}-\hat{y}_{uj}),
    \end{equation}
    where $\mathcal{N}_{u}$ denotes the set of items that interact with the user $u$,  
    $i\in\mathcal{N}_{u}$ is the positive item and $j\notin\mathcal{N}_{u}$ is the negative item sampled from unobserved interactions, $\sigma(\cdot)$ denotes the sigmoid function. 

    Secondly, during the LLM fine-tuning stage, we limit the classification space to match the actual item count and use cross-entropy as the loss function to mitigate the ‘hallucination’ issue, where items not present in the dataset are generated: 
    \begin{equation}
        \mathcal{L}_{\mathrm{LLM\_PEFT}} = -\sum_{n=1}^N y_n \log(p_n),
    \end{equation}
    where $y_n$ represents the ground truth label, and $p_n$ denotes the predicted probability for the $n$-th item interacting with the user. 
    Our fine-tuning thus optimizes the LLM to become a recommender, guided by the graph and sequential information.

\section{Theoretical Analysis of Complexity}
To elucidate the computational efficiency of our proposed hypergraph-based framework, we provide a detailed theoretical analysis encompassing both the hypergraph construction and convolutional propagation stages. For clarity and generality, we take the item-to-item (I2I) hypergraph as a representative case, as it captures the typical complexity patterns of other hypergraph variants (e.g., user-to-user). This analysis quantifies both time and space complexities, offering insights into the scalability and practicality of the proposed design. 
\subsection{Hypergraph Construction Complexity}
The total construction time consists of three stages: multimodal pairwise similarity computation $O(N^2 \cdot d)$, top-K neighbor selection $O(N\cdot \log k)$, and normalization over non-zero entries $O(N\cdot k)$. Thus, the total time complexity is:
\begin{equation}
        O(N^2 \cdot d+N\cdot \log k + N\cdot k) \approx O(N^2 \cdot d)
\end{equation}
The construction of the initial similarity matrix incurs a space complexity of $\mathcal{O}(N^2)$ due to its dense nature. However, following $k$-nearest neighbor (KNN) pruning, the matrix is sparsified, reducing the space requirement to $\mathcal{O}(N \cdot k)$, which significantly improves memory efficiency. 

\subsection{Hypergraph Convolution Complexity}
To enable efficient computation, we precompute the sparse matrix $\overline{\mathbf{H}}_{\iti} = \overline{\mathbf{I}} \cdot \overline{\mathbf{I}}^\top$, where each row has at most $k$ non-zero entries due to the KNN pruning, resulting in a practical storage size of $O(N \cdot k)$. This design transforms the hypergraph convolution into a single sparse-dense matrix multiplication $\overline{\mathbf{H}}_{\iti} \mathbf{X}_{\itt}$, with time complexity $\mathcal{O}(N \cdot k \cdot d)$ and space complexity $\mathcal{O}(N \cdot k + N \cdot d)$. This matches the complexity of standard GCN or HGNN layers while reducing redundant computation via precomputation. 

\subsection{User-to-User (U2U)  Hypergraph Complexity}

The complexity of U2U hypergraph construction and convolution is analogous to I2I hypergraph, with construction time complexity $O(M^2 \cdot N)$ replace $O(N^2 \cdot d)$. 
The resulting hypergraph matrix $\overline{\mathbf{H}}_{\text{U2U}} = \mathbf{A}\mathbf{A}^\top$ remains sparse and supports a similar convolution process. In this setting, the sparsity $k$ reflects the average number of user–item interactions per user. 
For simplicity, we omit the constant scale factors such as the number of modalities and layers, which would linearly scale the total complexity. 
The computational complexity and experimental details for hypergraph construction and convolution are provided in Sec.~\ref{subsec:cost}.
\section{EXPERIMENTS}

\subsection{Experimental Settings}

\subsubsection{Datasets}

    To evaluate our proposed method, we conduct experiments on three public benchmark datasets from Amazon \footnote{Datasets are available at http://jmcauley.ucsd.edu/data/amazon/links.html}\cite{mcauley2015image}: Sports (Sports and Outdoors), Beauty, and Toys (Toys and Games) with dataset statistics in Table \ref{Table:dataset}. 
    Each dataset includes visual and textual modalities. For the textual modality, we concatenate titles, brands, categories, and descriptions, processing them through BERT \footnote{https://huggingface.co/google-bert/bert-base-uncased} \cite{DBLP:journals/corr/abs-1810-04805} to obtain 768-dimensional features. For the visual modality, images are sourced from the original URLs, with 512-dimensional features extracted via VIT\footnote{https://huggingface.co/openai/clip-vit-base-patch32} \cite{radford2021learning}. 
    Following prior work \cite{kang2018self}, we split each user sequence by using the last interaction as the test set, the penultimate as the validation set, and all preceding interactions as the training set. 
\begin{table}[htbp]
    \centering
    \renewcommand{\arraystretch}{1.2}
    \caption{Statistics of the datasets. }
    \scalebox{1}{\begin{tabular}{c | c | c | c | c}
        \hline
        \textbf{Dataset} & \textbf{Users} & \textbf{Items} & \textbf{Interactions} & \textbf{Sparsity}\\
        \hline
        Sports & 35,598 & 18,357 & 260,739 & 99.960\% \\
        Beauty & 22,363 & 12,101 & 176,139 & 99.935\%  \\
        Toys & 19,412 & 11,924 & 148,185 & 99.936\%  \\
        \hline
    \end{tabular}}
    \label{Table:dataset}
\end{table}

    \subsubsection{Metrics} We evaluate top-K recommendation accuracy using three standard metrics across all datasets: Recall@K (R@K) and Normalized Discounted Cumulative Gain (N@K), with K set to 10 and 20. We report R@10, N@10, R@20, and N@20 across all models.


\subsubsection{Baseline methods} 
\begin{table*}[!ht]
        \caption{Performance of all comparison methods across three datasets, with the best results highlighted in bold and the second-best results underlined.}
	\label{Table:2}
	\renewcommand\arraystretch{1.25}
	\begin{center}
		{
		\begin{threeparttable}{
                \scalebox{1}{
			\begin{tabular}{l | c c | c c | c c | c c | c c | c c }
                    \hline
                    \multirow{2}{*}{\textbf{Baseline}} & \multicolumn{4}{c|}{\textbf{Sports}} & \multicolumn{4}{c|}{\textbf{Beauty}} & \multicolumn{4}{c}{\textbf{Toys}} \\
				\cmidrule(lr){2-5}\cmidrule(lr){6-9}\cmidrule(lr){10-13}
                    & \textbf{R@10} & \textbf{N@10} & \textbf{R@20} & \textbf{N@20} & \textbf{R@10} & \textbf{N@10} & \textbf{R@20} & \textbf{N@20} & \textbf{R@10} & \textbf{N@10} & \textbf{R@20} & \textbf{N@20} \\ 
                    \hline

                    \multicolumn{13}{c}{Conventional Recommendation Methods} \\
                    \hline
                    \textbf{LightGCN} \cite{he2020lightgcn}	& 0.0232 & 0.0120 & 0.0411 & 0.0165  & 0.0426  & 0.0208 & 0.0669  & 0.0269 & 0.0273  & 0.0138 & 0.0409  & 0.0172\\
                    \textbf{VBPR} \cite{he2016vbpr} & 0.0298&  0.0150 & 0.0485 & 0.0197  & 0.0436  & 0.0216 & 0.0724  & 0.0289  & 0.0376  & 0.0196 & 0.0565  & 0.0244\\
                    \textbf{GRCN} \cite{wei2020graph} & 0.0303 &  0.0152 & 0.0502 & 0.0202 & 0.0454  & 0.0225 & 0.0758 & 0.0305  & 0.0411  & 0.0214 & 0.0610  & 0.0264 \\
                    \textbf{LATTICE} \cite{zhang2021mining} & 0.0310 & 0.0155 & 0.0506 & 0.0204  & 0.0468  & 0.0234 & 0.0760  & 0.0307 & 0.0459  & 0.0228 & 0.0683  & 0.0284 \\
                    \textbf{SLMRec} \cite{tao2022self}& 0.0312 & 0.0156 & 0.0507 & 0.0205  & 0.0458  & 0.0227 & 0.0754  & 0.0301 & 0.0436  & 0.0223 & 0.0675  & 0.0284 \\
                    \textbf{BM3} \cite{zhou2023bootstrap} & 0.0316 & 0.0160 & 0.0517 & 0.0210   & 0.0446 &  0.0217 & 0.0722  & 0.0286 & 0.0402  & 0.0204 & 0.0618  & 0.0258\\
                    \textbf{FREEDOM} \cite{zhou2023tale} & 0.0314 & 0.0157 & 0.0507 & 0.0205 & 0.0516  & 0.0248 & 0.0805  & 0.0321  & 0.0467  & 0.0232 & 0.0700 & 0.0291\\
                    \textbf{LGMRec} \cite{guo2024lgmrec} & 0.0321 & 0.0164 & 0.0519 & 0.0213 & 0.0495  & 0.0240 & 0.0795  & 0.0316  & 0.0451  & 0.0228 & 0.0715 & 0.0294\\
                    \textbf{PGL} \cite{yu2025mind} & 0.0302 & 0.0152 & 0.0486 & 0.0198 & 0.0449  & 0.0220 & 0.0719  & 0.0288  & 0.0436  & 0.0212 & 0.0651 & 0.0266\\
                    \hline
                    \multicolumn{13}{c}{Large Language Models-based Methods} \\

                    \hline
                    
                    \textbf{Collm} \cite{zhang2023collm} & 0.0349 & 0.0182 & 0.0525 & 0.0226 & \underline{0.0681} &  \textbf{0.0393} & \underline{0.0960} & \underline{0.0463} & 0.0652  & \underline{0.0376} & 0.0905  & \underline{0.0439}  \\
                    \textbf{LLaRA} \cite{liao2024llara} & \underline{0.0367} & \underline{0.0203} & \underline{0.0562} & \underline{0.0252}  & 0.0667 &  0.0380 & 0.0925 & 0.0445 & \underline{0.0683}  & 0.0371 & \underline{0.0944}  & 0.0437\\
                    \hline
                    $\textbf{HeLLM}_{\textbf{pretrain}}$ (Ours) &  0.0329 & 0.0168 & 0.0542 & 0.0221  & 0.0527 &  0.0269 & 0.0820 & 0.0342 & 0.0498  &  0.0246 &  0.0786  & 0.0319 \\
                    \textbf{HeLLM} (Ours)&  \textbf{0.0412} & \textbf{0.0227} & \textbf{0.0617} & \textbf{0.0279}  & \textbf{0.0696} &  \underline{0.0384} & \textbf{0.1021} & \textbf{0.0465} & \textbf{0.0751}  &  \textbf{0.0404} &  \textbf{0.1081}  & \textbf{0.0488} \\
                \hline
		      \end{tabular}}  
                }
		\end{threeparttable}}
	\end{center}
\end{table*}
To validate our model’s effectiveness, we compare it with various state-of-the-art recommender models, including i) the general collaborative filtering methods LightGCN \cite{he2020lightgcn}, ii) multimodal GNN-based CF methods (e.g., VBPR \cite{he2016vbpr}, GRCN \cite{wei2020graph}, LATTICE \cite{zhang2021mining}, SLMRec \cite{tao2022self}, BM3 \cite{zhou2023bootstrap},  FREEDOM \cite{zhou2023tale}, LGMRec \cite{guo2024lgmrec}, PGL \cite{yu2025mind}) and iii) LLM-based methods Collm \cite{zhang2023collm} and LLaRA \cite{liao2024llara}. 
The detailed descriptions of each method are provided below.
\begin{itemize}
    \item \textbf{LightGCN} \cite{he2020lightgcn} removes feature transforms and nonlinearity from GCNs, relying solely on linear neighborhood aggregation and layer combination for recommendation. 
    \item \textbf{VBPR} \cite{he2016vbpr} firstly integrates visual features into a scalable factorization model to uncover users' visual preferences. This pioneering work established the foundation for multimodal recommendation. 
    \item \textbf{GRCN} \cite{wei2020graph} adaptively prunes false-positive edges in interaction graphs via refining layers, then applies graph convolution. 
    \item \textbf{LATTICE}  \cite{zhang2021mining} mines latent item-item structures from multimodal content by learning modality-aware item graphs to enhance item representation. 
    \item \textbf{SLMRec} \cite{tao2022self}  introduces self-supervised learning by augmenting multimodal content through feature dropout and masking to enhance item representations. 
    \item \textbf{BM3}  \cite{zhou2023bootstrap} bootstraps latent contrastive views via dropout and jointly optimizes interaction reconstruction and cross-/intra-modal alignment to learn representations. 
    \item \textbf{FREEDOM} \cite{zhou2023tale} simplifies LATTICE by fixing item-item graphs and denoising the user-item interaction graph via degree-sensitive edge pruning. 
    \item \textbf{LGMRec} \cite{guo2024lgmrec} decouples collaborative and multimodal signals via local graphs and leverages a hypergraph module to model global user interests. 
    \item \textbf{PGL} \cite{yu2025mind} extracts the principal subgraphs with global and local aware operators to capture both co-occurrence patterns and personalized signals.
    \item \textbf{Collm}  \cite{zhang2023collm} enhances LLM-based recommendation by treating collaborative signals as an independent modality and aligning them with LLM inputs via a learnable mapping module, while using LoRA for fine-tuning. 
    \item \textbf{LLaRA} \cite{liao2024llara} enhances sequential recommendation by aligning ID-based behavioral embeddings with the LLM input space through hybrid prompting. 
    \end{itemize}
To ensure a fair comparison, all LLM-based models adopt the same pre-trained backbone, LLaMA-2 7B \cite{touvron2023llama}, and share the same textual prompt templates as illustrated in Fig.\ref{fig:framework}. Specifically, CoLLM and LLaRA incorporate item embeddings derived from LightGCN \cite{he2020lightgcn} and SASRec \cite{kang2018self}, respectively. All LLM methods are optimized using the Cross-Entropy loss. The results are summarized in Table \ref{Table:2}. 

\addtocounter{MaxMatrixCols}{10}

\subsubsection{Implementation details}
    We implemented all the experiments using PyTorch on an RTX 4090 GPU with 24GB memory. 
    For a fair comparison with previous works ~\cite{zhou2023tale,zhou2023bootstrap}, the recommender dimension $d$ is set to 64 for all pre-trained models. 
    Our model is trained using the Adam optimizer \cite{kingma2014adam}, with learning rates and batch sizes set at 0.0001 and 1024 for pre-training, and 0.0003 and 7 for fine-tuning, respectively. 
    We use the Xavier initializer \cite{glorot2010understanding} for initializing the user and item embeddings. 
    The hyperparameters $\alpha$ and $\beta$ are optimized by searching from 0.1 to 0.9 in 0.2 increments. 
    The number of k-nearest neighbors varies from 1 to 30 in increments of 5. 
    The contrastive learning temperature $\tau$ is adjusted from 0.1 to 1 in 0.1 increments. 
    More details are presented in Sec. \ref{subsec:example}.
    
    \subsection{Performance Comparison}

The experimental results, presented in Table \ref{Table:2}, reveal the following key observations:

\begin{itemize}
    \item \textbf{Overall Superiority of HeLLM:} Our proposed HeLLM consistently achieves the best or near-best performance across all metrics and datasets. It outperforms the strongest baseline in most cases, often by a notable margin. Even in cases where a baseline is very competitive, HeLLM remains on par or better. This demonstrates the effectiveness and robustness of our approach across different domains. The gains validate that augmenting LLMs with hypergraph-based context and sequential information indeed leads to better recommendation quality. By integrating hypergraphs and time-series signals, HeLLM identifies latent shared user interests and infers multimodal item relationships, thereby enhancing the precision of its context-aware predictions through dynamically refined representations.

    \item \textbf{Impact of Multimodal Information:} The results show that methods utilizing multimodal features (e.g. GRCN, LATTICE, SLMRec, BM3) significantly outperform the pure CF model LightGCN on all datasets. This confirms that incorporating visual and textual information facilitates learning more comprehensive representations and enables a deeper understanding of diverse user preferences, thereby promoting more personalized and context-aware recommendations. 
    Inheriting the multimodal feature integration framework of the I2I hypergraph, our approach extends this foundation by augmenting relational reasoning capabilities within the LLM architecture. 

    \item \textbf{LLM-based vs GNN-based: }Comparing the LLM-augmented models (HeLLM, CoLLM, LLaRA) with the pure GNN-based models (e.g., GRCN, FREEDOM), it shows that LLM-based approaches have distinct advantages on these datasets. In particular, they excel in scenarios with sparse user-item interactions. For example, on the three sparse datasets (more than 99.9\% sparsity), CoLLM and LLaRA already outperform GRCN and FREEDOM, and HeLLM further improves on CoLLM/LLaRA. The LLM’s vast pre-trained knowledge and language reasoning enable to infer user preferences even when data is limited, by making associations or generalizations that a pure CF model might miss. Moreover, the flexibility of LLMs to adapt with minimal data (thanks to PEFT) enables effective fine-tuning across diverse domains. This indicates that LLM-based recommenders can generalize better and cope with cold-start situations, as they don’t rely solely on dense graph connectivity but can utilize semantic cues.

\begin{table*}[!ht]
\centering
\caption{Resource Consumption Across All Stages of Our Framework on Three Benchmark Datasets}
\label{tab:cost}
\renewcommand{\arraystretch}{1.2}
\setlength{\tabcolsep}{1.3mm}
    \scalebox{0.83}{
\begin{tabular}{l|ccc|cc|cc|cc|cc|ccc}
\toprule
\textbf{Dataset} & \textbf{Users} & \textbf{Items} & \textbf{Sparsity} & \textbf{U-I Time} & \textbf{U-I Mem} & \textbf{U2U Time} & \textbf{U2U Mem} & \textbf{I2I Time} & \textbf{I2I Mem} & \textbf{Pretrain Time} & \textbf{Pretrain Mem} & \textbf{FT Trainable} & \textbf{FT Time} & \textbf{FT Mem} \\
\midrule
Sports & 35,598 & 18,357 & 99.96\% & 281.6s & 8.59MB & 53.6s & 195.69MB & 54.2s & 22.92MB & 2h57min & 16,151MB & 1.66\% & 15h33min & 22,227MB \\
\midrule
Beauty & 22,363 & 12,101 & 99.94\% & 80.3s & 5.87MB & 14.0s & 101.67MB & 33.7s & 139.53MB & 41min & 7,515MB & 1.66\% & 9h49min & 22,537MB \\
\midrule
Toys   & 19,412 & 11,924 & 99.94\% & 58.5s & 4.91MB & 10.2s & 57.48MB & 33.1s & 79.53MB & 3h38min & 6,633MB & 1.66\% & 8h6min & 22,047MB \\
\bottomrule
\end{tabular}}
\end{table*}

    \item  \textbf{Combining Graph and Sequence is Beneficial:} 
    We empirically validate the core hypothesis that both graph-structured collaborative signals and sequential user behaviors are indispensable for effective recommendation. Our proposed HeLLM -which jointly models user-item interactions via hypergraph-based global context and sequential dynamics—consistently outperforms CoLLM (collaborative only) and LLaRA (sequential only) across all benchmarks. While CoLLM (based on LightGCN) and LLaRA (based on SASRec) yield statistically comparable performance, neither achieves consistent dominance, indicating the limitations of using either global or local information in isolation. In contrast, HeLLM captures short-term personalized trends through sequential modeling and higher-order cross-user/item semantics through hypergraph structure, yielding a more comprehensive representation of user preferences. The superior performance of HeLLM highlights the complementary nature of graph and sequential signals, demonstrating that their integration leads to a deeper understanding of user behavior and significantly enhances recommendation quality.
\end{itemize}

\begin{table}[!tp]
    \caption{Performance comparison between HeLLM and its variants.}
    \centering
    \small
    \renewcommand\arraystretch{1.25}
    \setlength{\tabcolsep}{1.3mm}
    \scalebox{0.83}{
    \begin{tabular}{l | c c | c c | c c }
        \hline
        \multirow{2}{*}{\textbf{}} & \multicolumn{2}{c|}{Sports} & \multicolumn{2}{c|}{Beauty} & \multicolumn{2}{c}{Toys}\\
        \cmidrule(lr){2-7}
         & R@20 & N@20 & R@20 & N@20 & R@20 & N@20 \\
        \hline
        $\textbf{HeLLM}_{\textbf{pretrain}}$ w/o-U2U & 0.0486 & 0.0203 & 0.0736 & 0.0294 & 0.0592 & 0.0247 \\
        $\textbf{HeLLM}_{\textbf{pretrain}}$ w/o-I2I  & 0.0495 & 0.0205 &  0.0719 & 0.0290 & 0.0647 & 0.0270 \\
        $\textbf{HeLLM}_{\textbf{pretrain}}$ w/o-CL  &  0.0510 & 0.0209 &  0.0774 & 0.0315 & 0.0681 & 0.0281 \\
        \hline
        $\textbf{HeLLM}$ w/o-GNN\_PEFT  &  0.0582 & 0.0272 &  0.1014 & 0.0440 & 0.1011 &  0.0436 \\
        $\textbf{HeLLM}$ w/o-GNN\_ID  &  0.0573 & 0.0256 &  0.0991 & 0.0465 & 0.0993 &  0.0434 \\
        $\textbf{HeLLM}$ w/o-Seq\_ID  &  0.0550 & 0.0232 &  0.0949 & 0.0428 & 0.0951 & 0.0412 \\
        \hline
        $\textbf{HeLLM}_{\textbf{pretrain}}$  & 0.0542 & 0.0221 & 0.0820 & 0.0342 & 0.0786 & 0.0319 \\
         \textbf{HeLLM} & \textbf{0.0617} & \textbf{0.0279} & \textbf{0.1021} & \textbf{0.0465} & \textbf{0.1081} & \textbf{0.0488} \\
        \hline
    \end{tabular}
    }
    \label{Table:3}
\end{table}
\subsection{Ablation Study}

\begin{figure*}[tbp]
\centering   
\setcounter{subfigure}{0} 
    \subfigure[$\text{Top-}K$]{

\includegraphics[width=4.1cm]{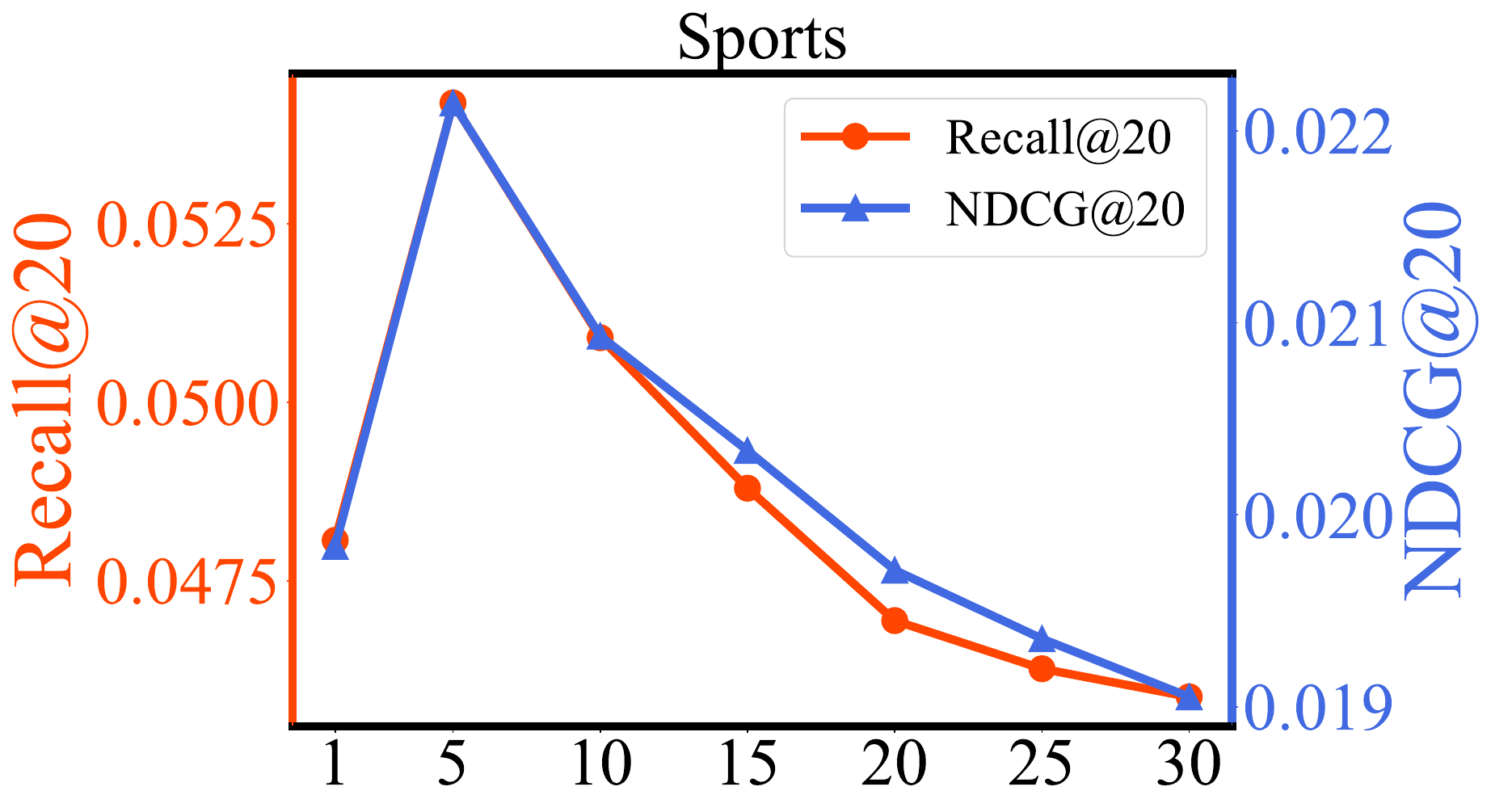}  

} 
\setcounter{subfigure}{3} 
\subfigure[CL Temperature $\tau$ ] {
\includegraphics[width=4.1cm]{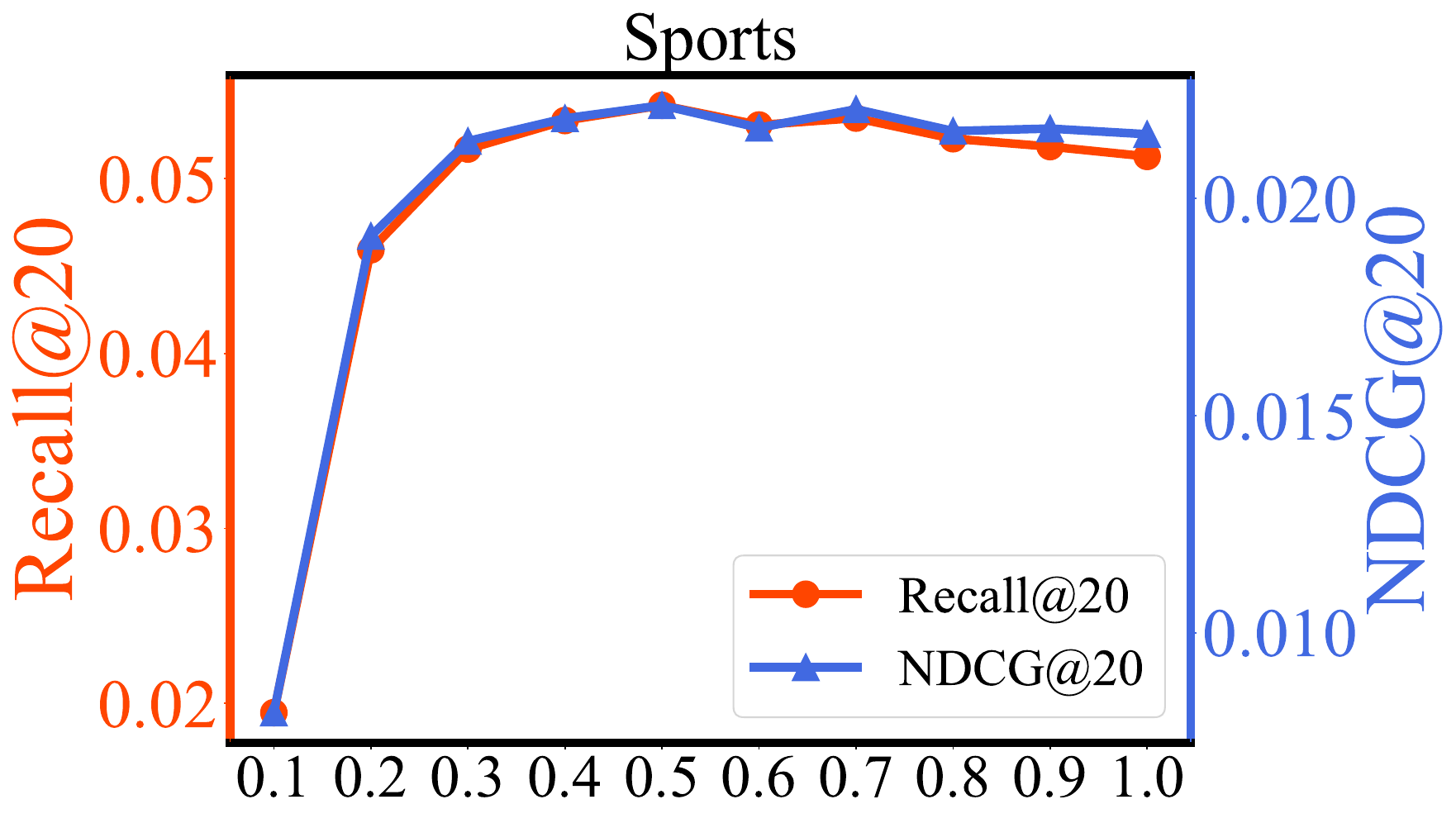}  
}     
\setcounter{subfigure}{6} 
\subfigure[ $L_{2}$ Regularization Coefficient $\lambda$] {
\includegraphics[width=4.1cm]{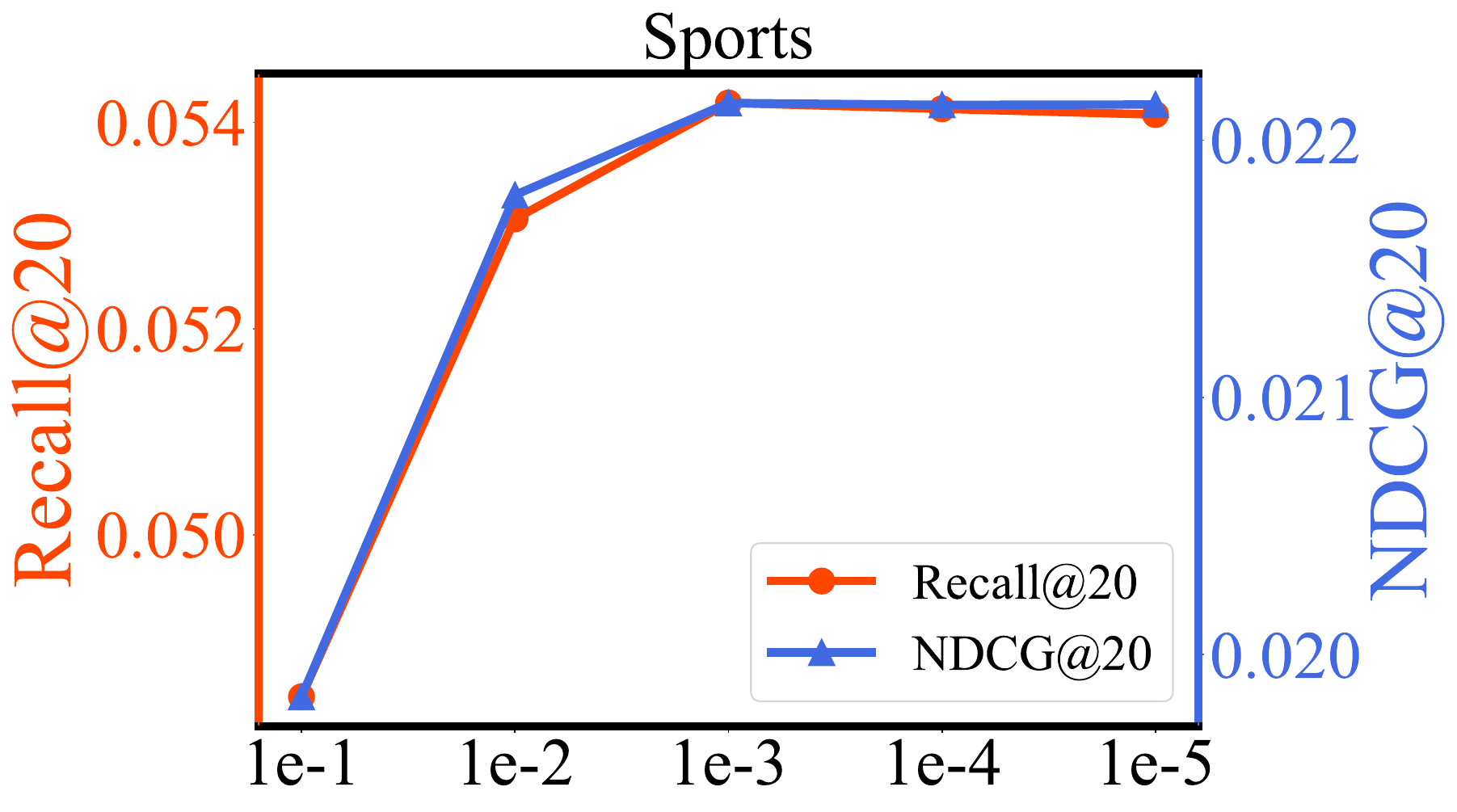}  
}     
\setcounter{subfigure}{9} 
\subfigure[Embedding Dimensionality $d$] {
\includegraphics[width=4.1cm]{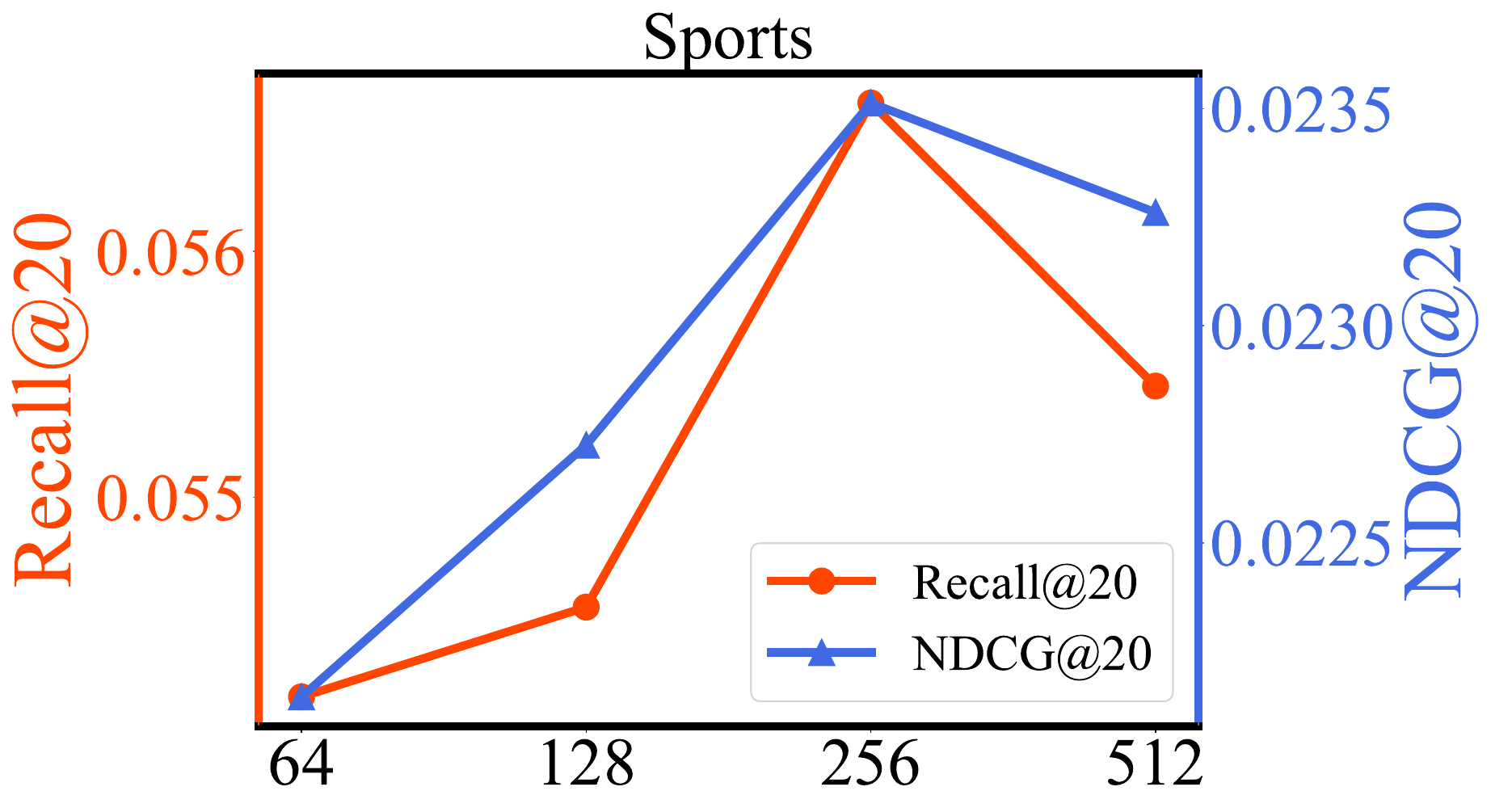}  
}     
\setcounter{subfigure}{1} 
\subfigure[$\text{Top-}K$] {
\includegraphics[width=4.1cm]{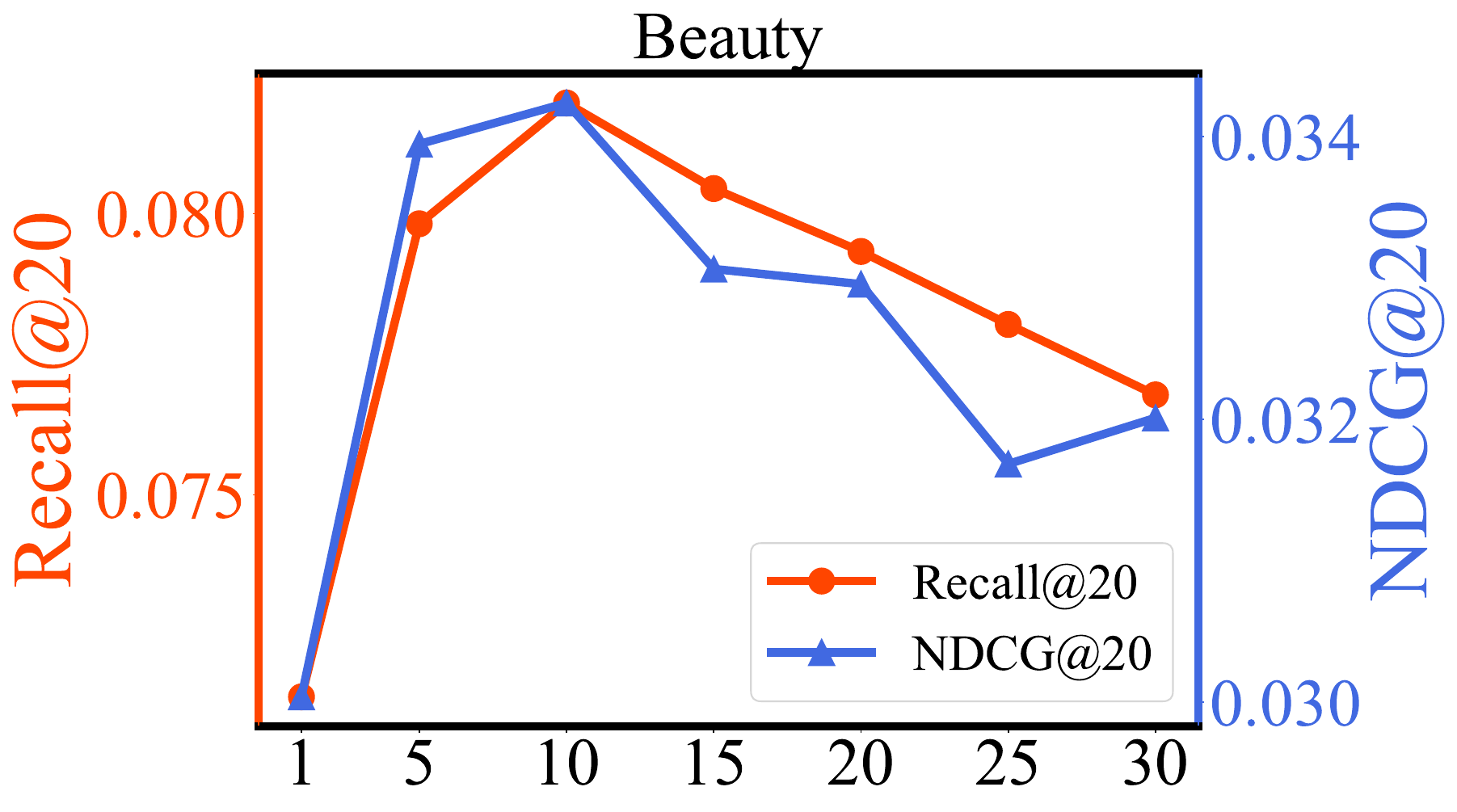}  
}     
\setcounter{subfigure}{4} 
\subfigure[CL Temperature $\tau$ ] {
\includegraphics[width=4.1cm]{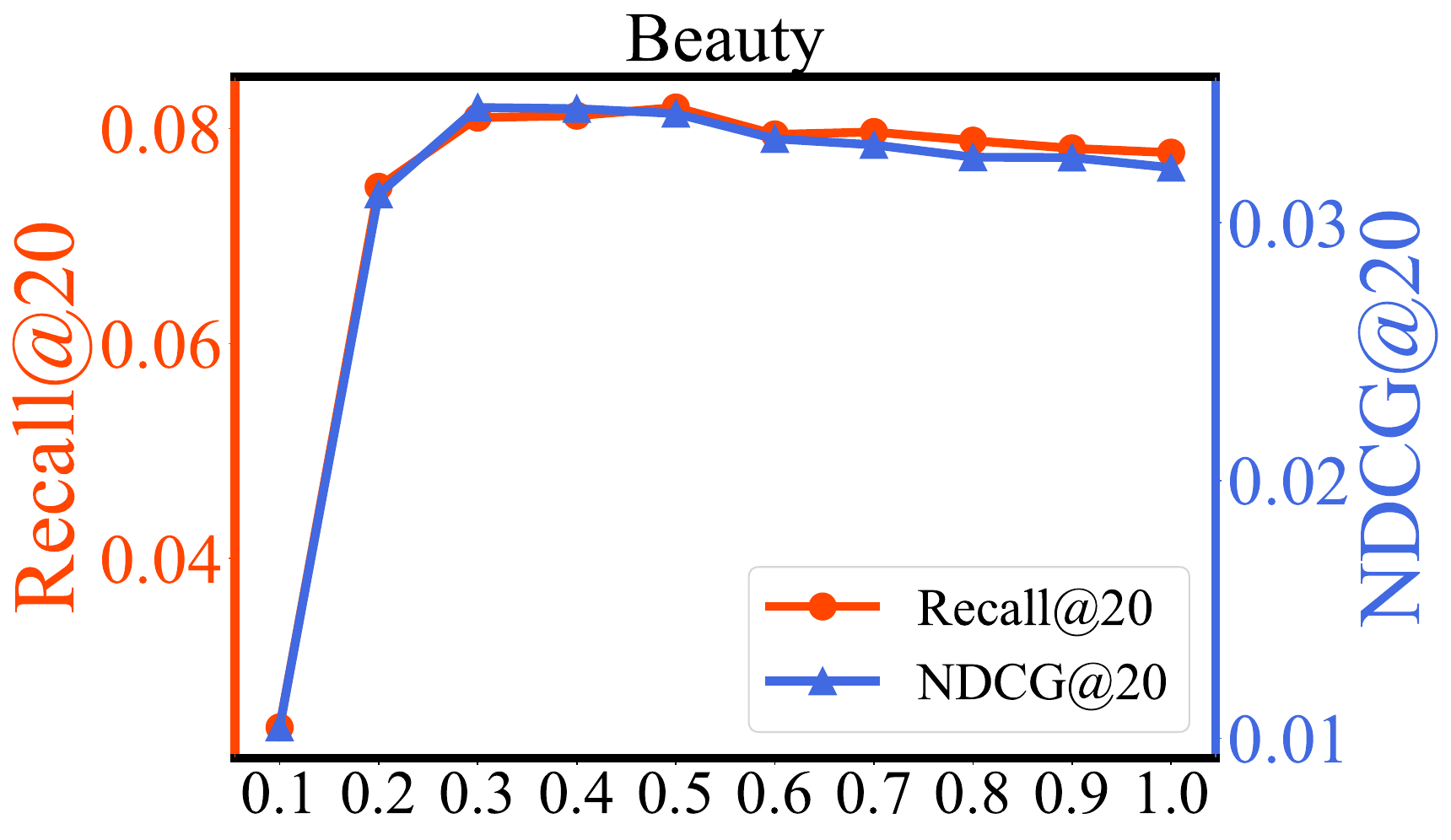}  
}     
\setcounter{subfigure}{7} 
\subfigure[$L_{2}$ Regularization Coefficient $\lambda$] {
\includegraphics[width=4.1cm]{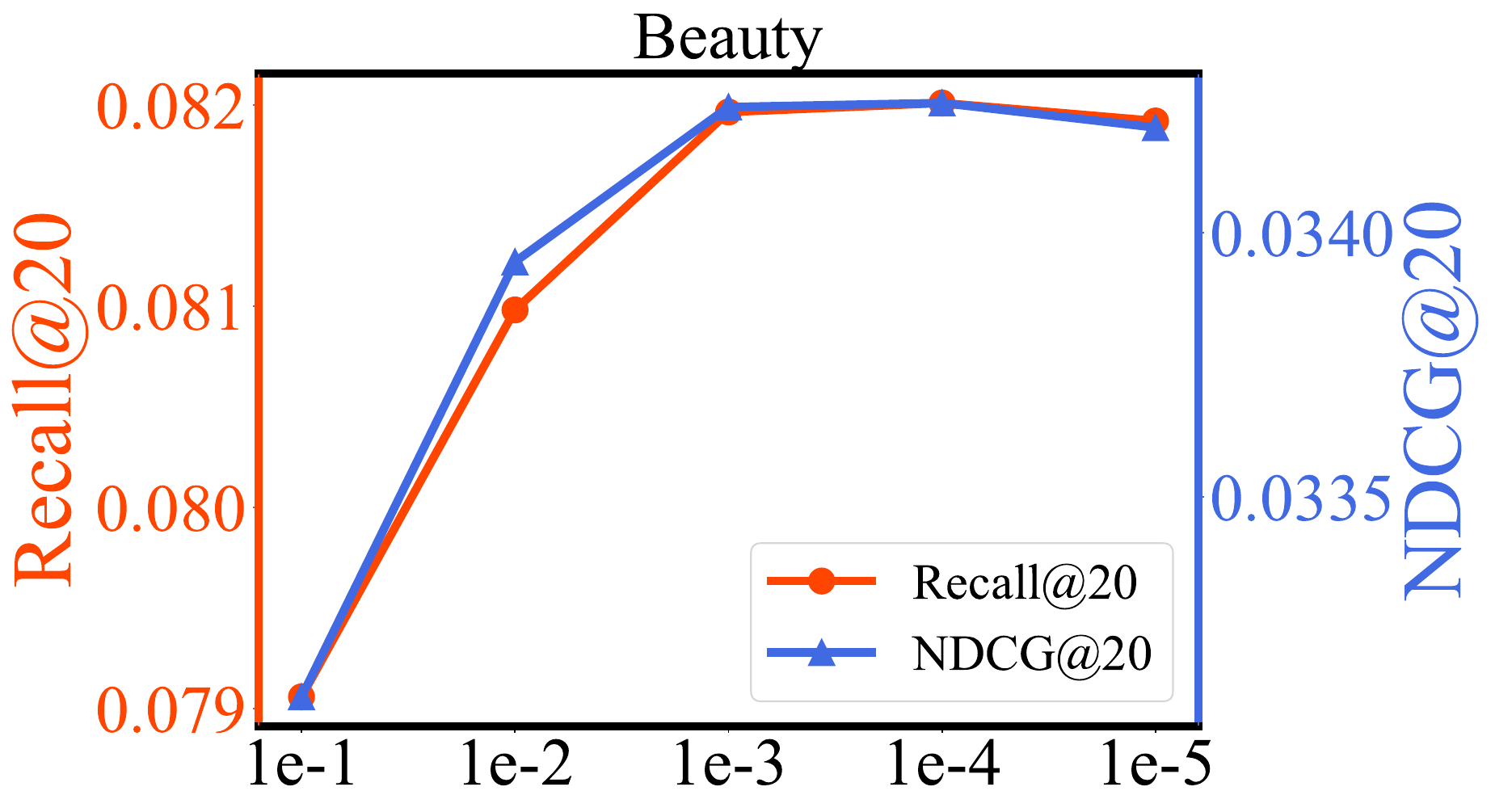}  
}     
\setcounter{subfigure}{10} 
\subfigure[Embedding Dimensionality $d$] {
\includegraphics[width=4.1cm]{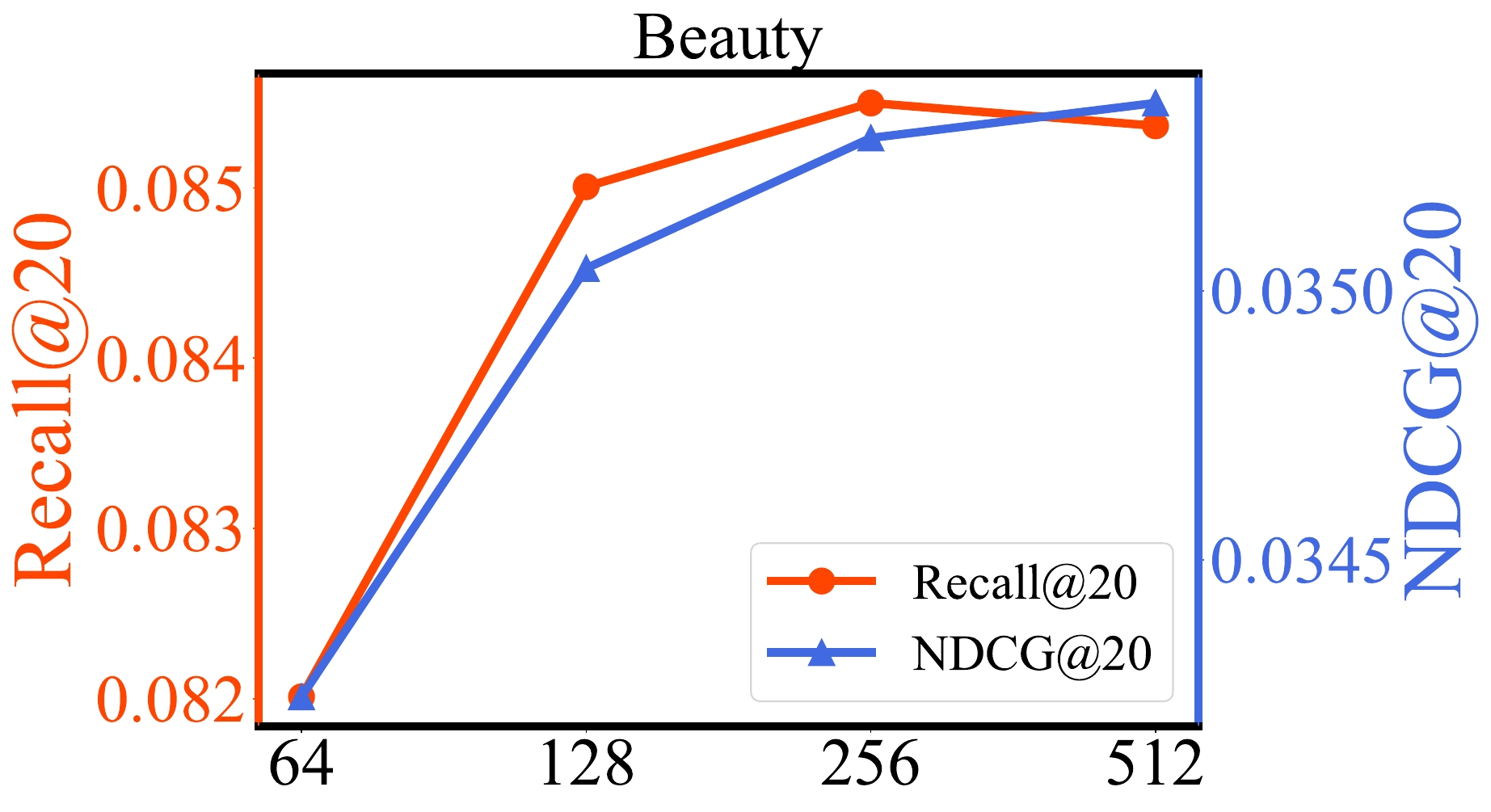}  
}     
\setcounter{subfigure}{2} 
\subfigure[$\text{Top-}K$] {
\includegraphics[width=4.1cm]{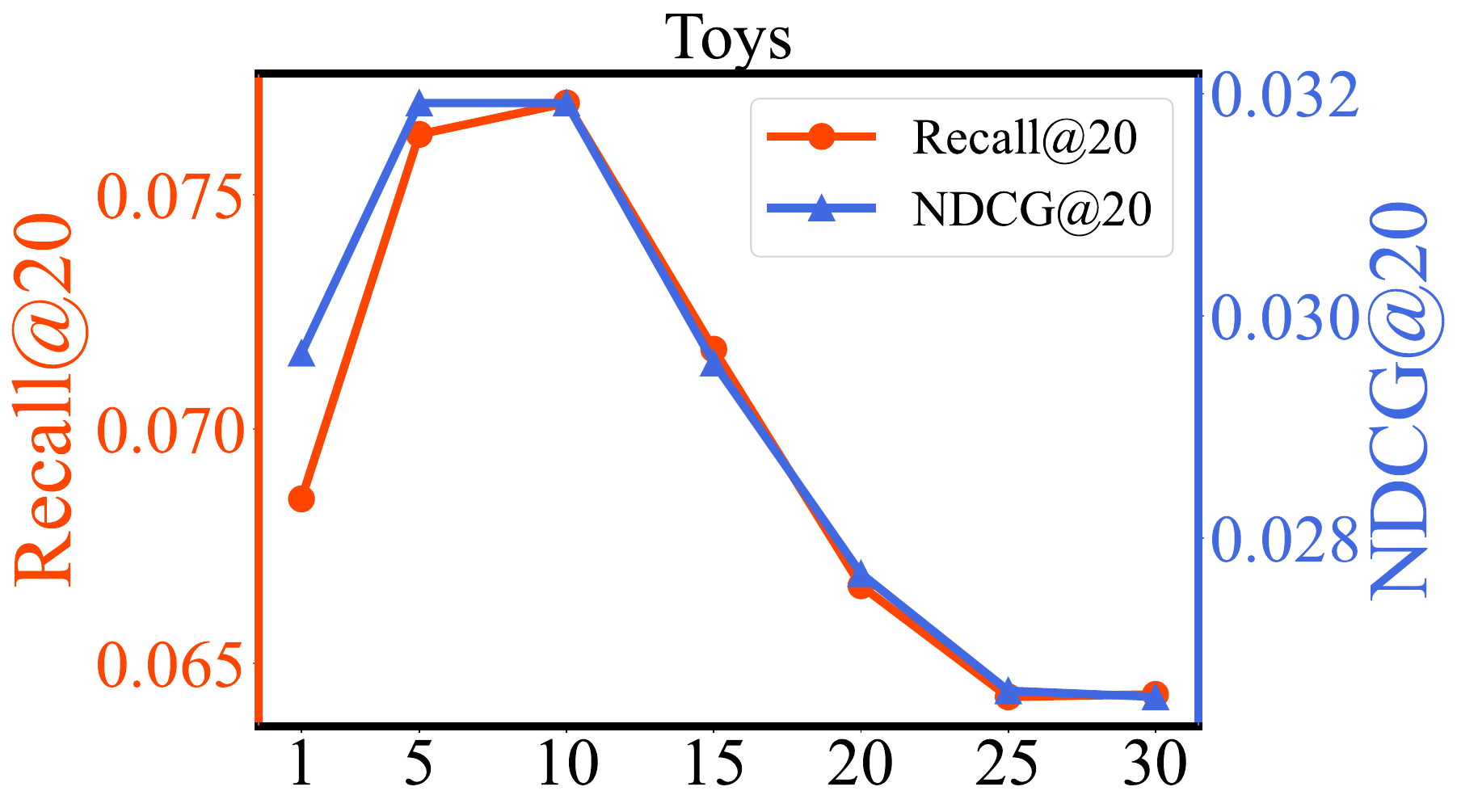}  
}     
\setcounter{subfigure}{5} 
\subfigure[CL Temperature $\tau$] {
\includegraphics[width=4.1cm]{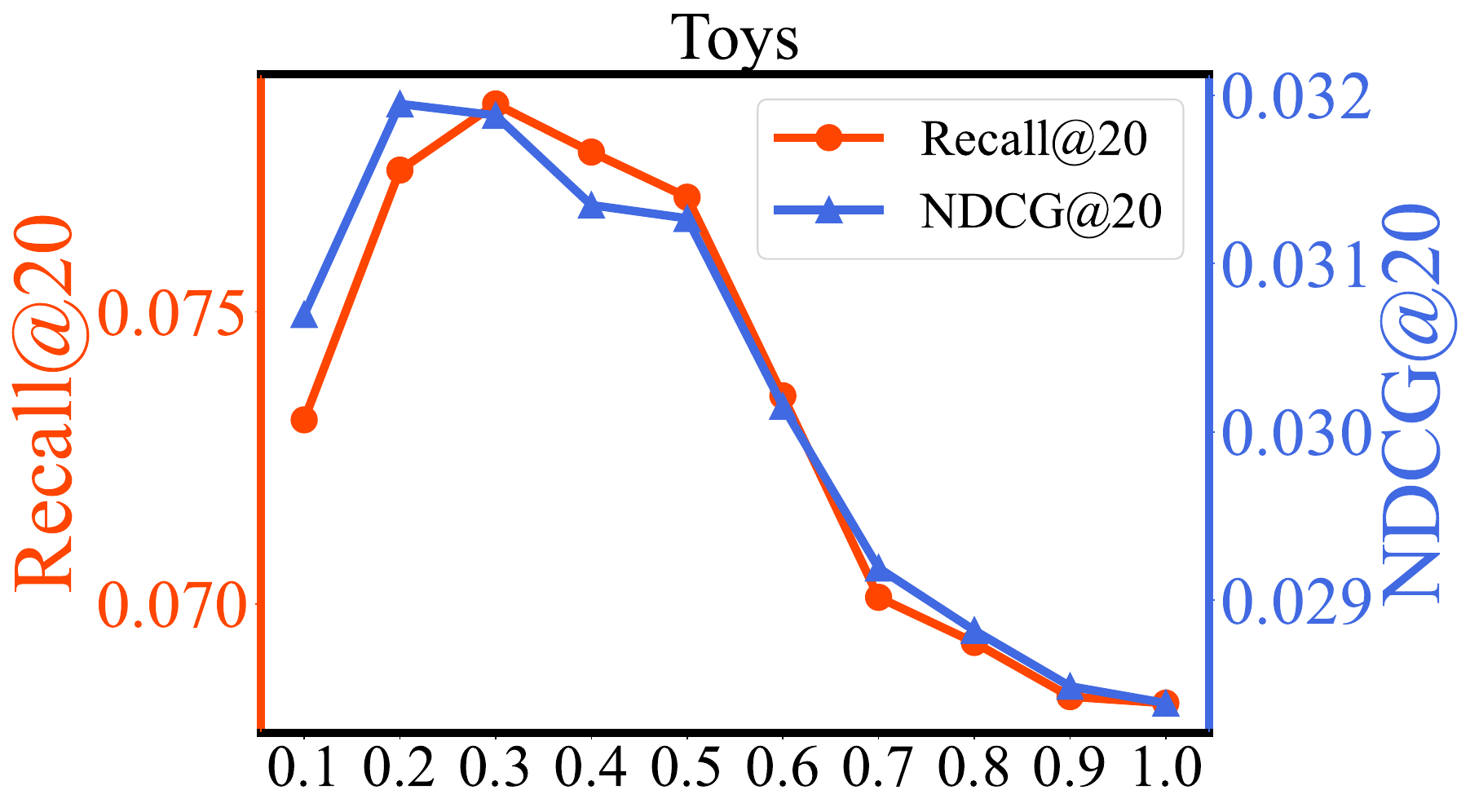}  
}     
\setcounter{subfigure}{8} 
\subfigure[$L_{2}$ Regularization Coefficient $\lambda$] {
\includegraphics[width=4.1cm]{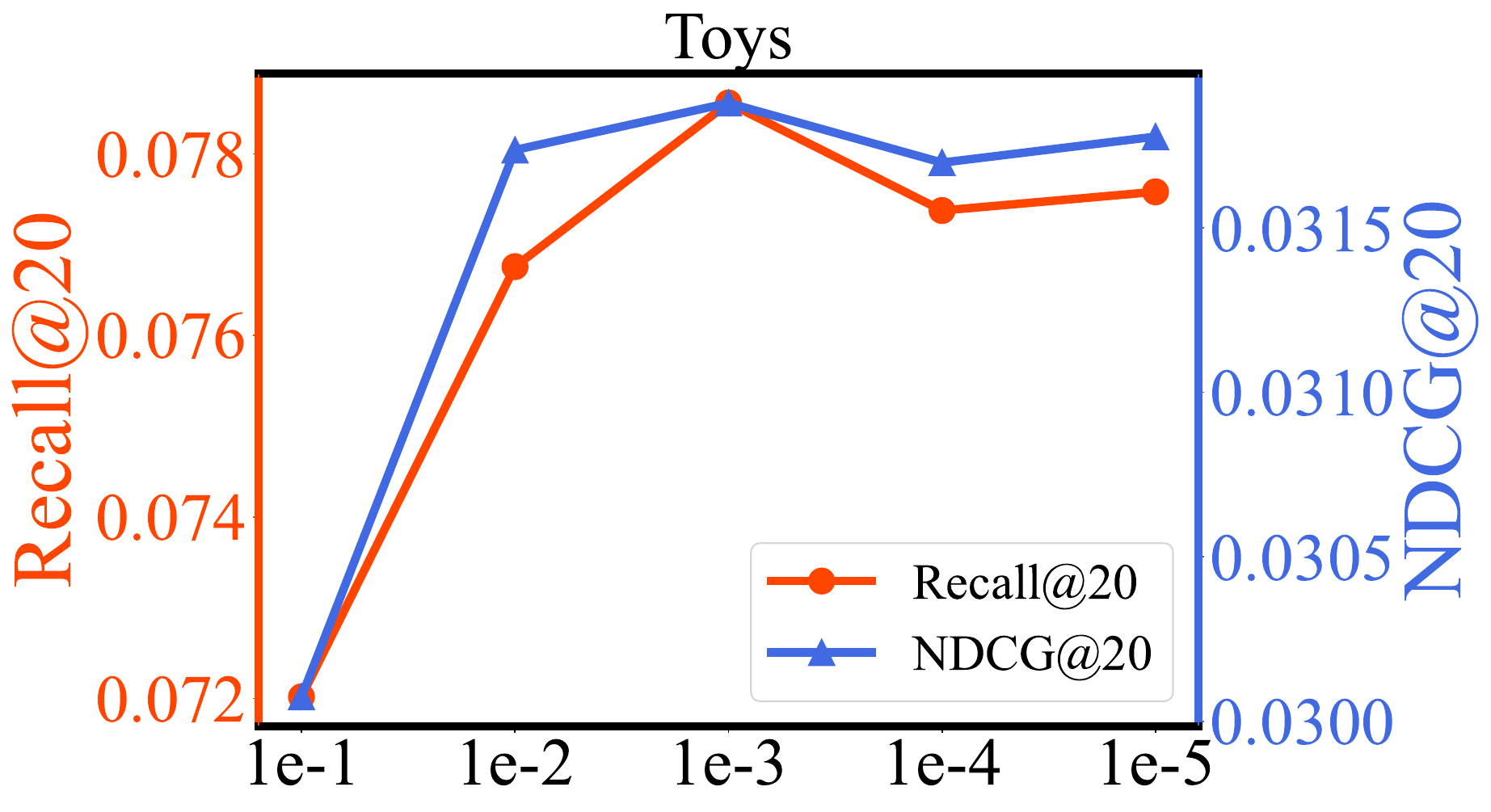}  
}     
\setcounter{subfigure}{11} 
\subfigure[Embedding Dimensionality $d$] {
\includegraphics[width=4.1cm]{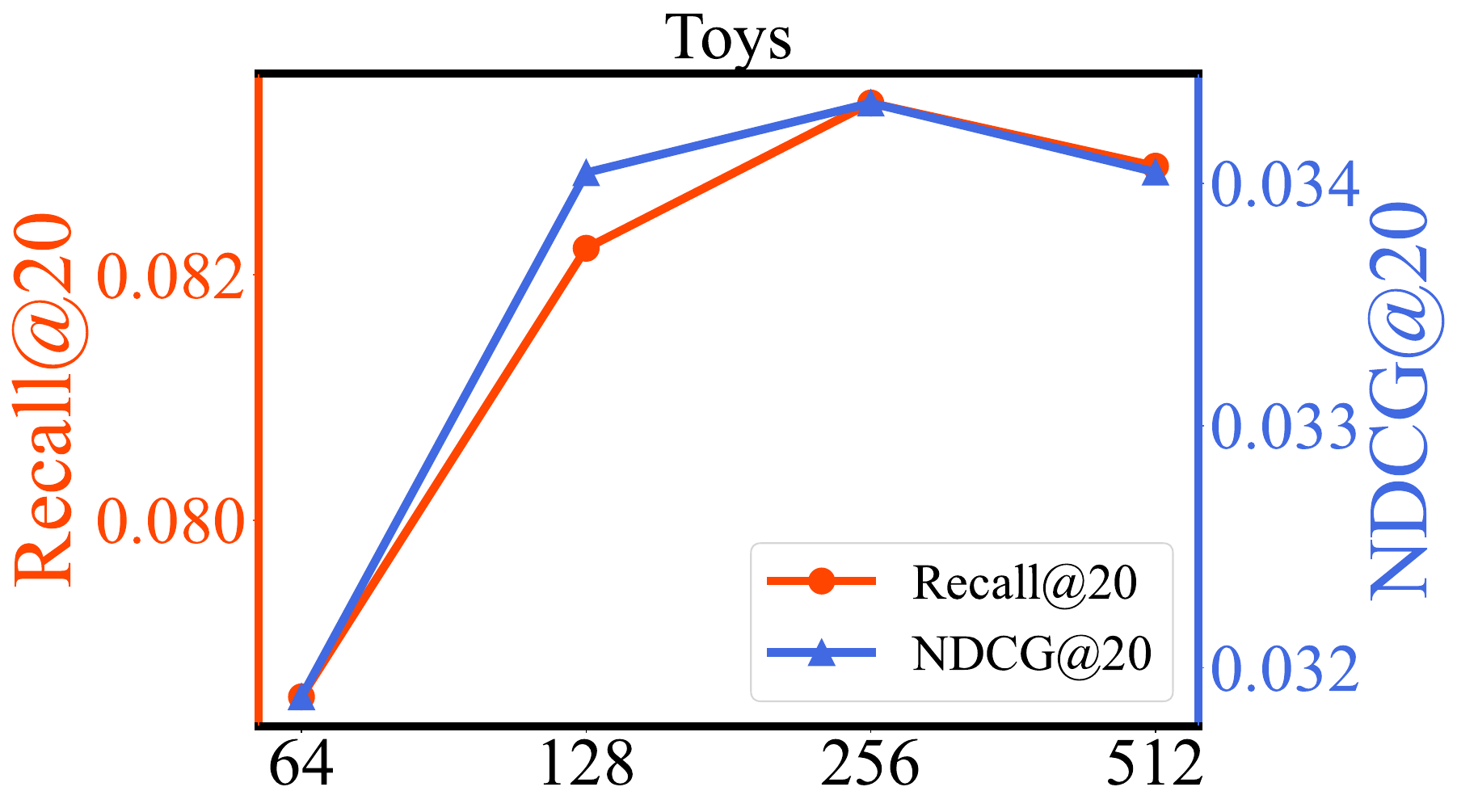}  
}     
\caption{Performance comparison w.r.t various hyperparameters on the three datasets.}
\label{hyperparameters}
\end{figure*}

To systematically evaluate the contribution of each component in HeLLM, we conduct a comprehensive ablation study, as summarized in Table \ref{Table:3}. Our analysis encompasses both the recommendation pre-training phase $\textbf{HeLLM}_{\textbf{pretrain}}$ and the LLM fine-tuning stage $\textbf{HeLLM}$, where key architectural elements are selectively removed to assess their individual impact. From this, we can observe that: 

\begin{figure}[tbp]
    \centering
    \includegraphics[width=0.23\textwidth]{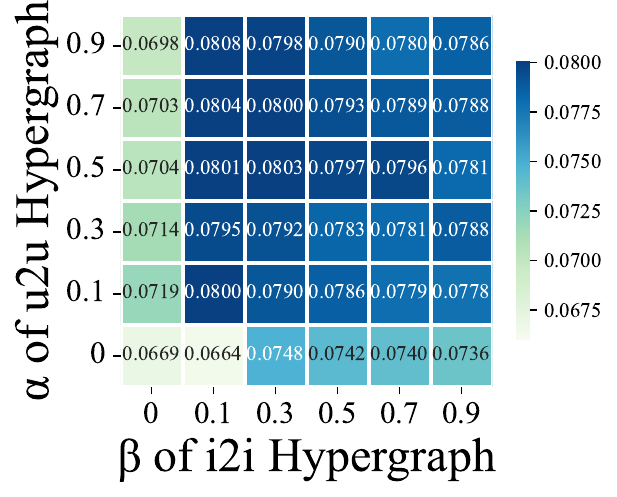}
    \includegraphics[width=0.23\textwidth]{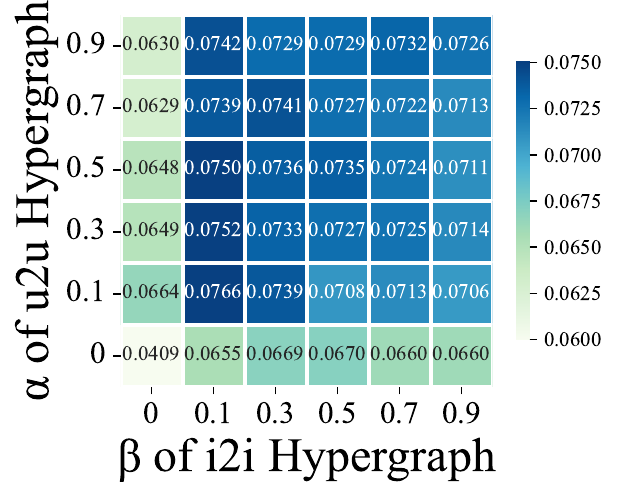}
    \caption{Performance w.r.t the effect of synergistic contrastive learning coefficients  $\alpha$ for u2u hypergraph and $\beta$ for i2i hypergraph on Beauty and Toys datasets (Recall@20). }
    \label{1.u2u_i2i_cl_ratio}
    \vspace{-5pt}
\end{figure}
    $\bullet$ \textbf{Effectiveness of $\textbf{HeLLM}_{\textbf{pretrain}}$ and Multi-View Hypergraph Learning.} We first conducted experiments for variants of the recommender pre-training $\textbf{HeLLM}_{\textbf{pretrain}}$ by removing different hypergraph modules and contrastive learning. The omission of the U2U hypergraph (w/o-U2U) leads to the most significant performance drop, particularly in the Sports and Toys datasets. This confirms the pivotal role of the U2U hypergraph in modeling complex shared preferences among users. Similarly, excluding I2I hypergraph (w/o-I2I) results in noticeable degradation, especially in the Beauty dataset, underscoring the importance of item-item multimodal semantic relations. The impact of contrastive learning (w/o-CL) is also evident, with a decline across all domains, reaffirming its role in enforcing robust representation learning. Notably,  our $\textbf{HeLLM}_{\textbf{pretrain}}$ consistently outperforms all GNN-based baselines in the pre-training phase, verifying its effectiveness in capturing multi-view graph topological information. These findings highlight the advantage of hypergraphs for capturing shared user preferences and intricate multimodal semantic correlation among items. 

  $\bullet$ \textbf{Contribution of Graph-Enhanced LLM Fine-Tuning.} In the fine-tuning phase, we assess the significance of graph-based augmentation. The removal of graph-enhanced PEFT (w/o-GNN\_PEFT) results in a severe performance drop, highlighting the necessity of graph-conditioned parameter-efficient fine-turning. Similarly, omitting GNN embeddings in the input (w/o-GNN\_ID) leads to a noticeable decline in ranking quality, suggesting that graph features serve as critical inductive biases to preserve structural information. Furthermore, the absence of sequential embeddings (w/o-Seq\_ID) negatively impacts performance, with an observable decline in the sequentially sensitive, verifying its essential role in preserving temporal and contextual consistency. 
  HeLLM integrates both GNN-based modeling and time-series analysis, surpassing alternative variants. This highlights the necessity of both components, as GNNs effectively capture structural dependencies while time-series models track temporal dynamics, jointly providing a holistic understanding of user behavior.

\subsection{Computation Cost and Efficiency}
\label{subsec:cost}
Table~\ref{tab:cost} provides a detailed overview of the computational cost and resource consumption of our framework across all pipeline stages. 
For each dataset, we report the number of users and items, interaction sparsity, and the time and memory required to construct the user-item (U-I) graph as well as the user-user (U2U) and item-item (I2I) hypergraphs. 
These graph and hypergraph construction processes benefit from our sparsity-aware design, which reduces the storage requirement from an initial dense $\mathcal{O}(N^2)$ matrix to $\mathcal{O}(N \cdot k)$ after KNN pruning, leading to significantly lower memory overhead (e.g., less than 200MB for U2U construction on the Sports dataset). 
All hypergraph matrices are precomputed once and reused throughout the training pipeline, effectively minimizing redundant computation and memory overhead.  

In addition, we further measure GPU-side resource requirements during the pretraining (PT) and fine-tuning (FT) phases, including wall-clock time, peak GPU memory usage, and the proportion of trainable parameters. 
During FT, only 1.66\% of parameters are updated with a batch size of 7, demonstrating that the proposed PEFT strategy enables efficient adaptation while keeping the memory footprint manageable. 
For example, our largest dataset, Sports (35,598 users and 18,357 items), can be processed within 3 hours of pretraining using 16GB GPU memory, and fine-tuning takes approximately 15 hours with 22GB GPU memory on a single NVIDIA RTX 4090 (24GB). 

\subsection{Impact Study of Hyperparameters}
\label{subsec:example}
    The model's performance is significantly dependent on the pre-training phase. To this end, we conduct a comprehensive analysis of the impact of various hyperparameters. 
    Consequently, we thoroughly investigate the effects of synergistic contrastive learning, k-nearest neighbors, contrastive learning temperature $\tau$, $L_{2}$ regularization coefficient, and the latent dimensionality. 
    
\subsubsection{Impact of Synergistic Contrastive Learning} 
As shown in Fig. \ref{1.u2u_i2i_cl_ratio}, $\alpha$ and $\beta$ 
 control the contribution of user-user (U2U) and item-item (I2I) hypergraph-based contrastive signals, respectively. 
 When both coefficients are set to zero, effectively disabling the contrastive learning modules, the model suffers a significant drop in performance, indicating that merely relying on structural encoders without contrastive regularization limits the model’s ability to learn discriminative representations. 
 Conversely, enabling both $\alpha$ and $\beta$ yields notable improvements, even with small values such as $\alpha = 0.1$ and $\beta = 0.1$, suggesting that introducing lightweight contrastive supervision already yields strong gains. 
This supports our hypothesis that the complementary signals captured by U2U and I2I views play a synergistic role in guiding pretraining. 
Moreover, the performance surface exhibits a relatively smooth basin centered around low to moderate $\alpha$ and $\beta$ , implying robustness to hyperparameter perturbation. However, excessively large values (e.g., $\alpha = 0.9$ and $\beta = 0.9$) do not yield further gains, and may even lead to slight degradation, likely due to over-regularization overriding structural cues. 



\subsubsection{Impact of Top-$\mathrm{K}$ }

As shown in Fig. \ref{hyperparameters}(a)-(c), it reveals that $\mathrm{K}=1$ is insufficient for capturing meaningful information, while $\mathrm{K}=5$ and $\mathrm{K}=10$ make an optimal balance. As $\mathrm{K}$ increases beyond this range, performance deteriorates due to the inclusion of irrelevant or noisy neighbors, which degrade the aggregation of relevant information in hypergraph message passing. Larger top-K values introduce redundancy, weakening the influence of key user-item relationships and ultimately reducing recommendation precision.

\subsubsection{Impact of Contrastive Learning (CL) Temperature Varied $\tau$ }
As shown in Fig. \ref{hyperparameters}(d)-(f), lower values of $\tau$ appear to overly separate positive samples from difficult negatives, disrupting the learned latent semantics.  However, for the Toys dataset, a lower $\tau$ appears beneficial, likely due to its higher proportion of challenging negative samples, which enables the model to capture richer representations. Conversely, excessive high value of $\tau$ reduces sample discrimination, diminishing model effectiveness. 

\subsubsection{Impact of $L_{2}$ Regularization Coefficient $\lambda$ }
As shown in Fig. \ref{hyperparameters}(g)-(i), the experiments reveal that $L_{2}$ regularization plays a pivotal role in balancing model performance by preventing overfitting while preserving essential feature information. Across the all three datasets, moderate values of $\lambda$ regularization, specifically around ${10}^{-3}$ consistently deliver optimal results in both metrics.

\subsubsection{Impact of Embedding Dimensionality $d$ }  


As shown in Fig. \ref{hyperparameters}(j)-(l), increasing the embedding dimensionality initially improves performance, with optimal results typically observed at d=256, after which it plateaus or slightly declines. This trend indicates that while higher dimensions enhance representational capacity, overly large embeddings introduce redundancy and unnecessary computational overhead. Thus, selecting an appropriate dimensionality is essential for balancing informativeness and efficiency.

    

    
\section{Conclusion}

In this study, we identify two key objectives for advancing LLM-based recommender systems: (1) enhancing the expressive power of pre-trained representations, and (2) effectively integrating rich graph structure information to strengthen the LLM’s ability to perceive relational patterns. To this end, we propose HeLLM, a novel framework that enhances LLMs’ graph reasoning capabilities for multimodal recommendation. During pre-training, user and item hypergraphs uncover shared interests and refine multimodal correlations, while a synergistic contrastive learning strategy enhances feature discriminability. In the fine-tuning phase, HeLLM employs mapped feature embeddings and graph pooling to inject enriched graph-structured and multimodal information into the LLM, strengthening its reasoning ability for recommendation tasks. Additionally, sequence-based embeddings capture temporal behavioral patterns, further enriching the model’s representation capacity.  Experiments on three benchmarks validate the effectiveness of our proposed method.


\vfill
\bibliographystyle{IEEEtran}
\bibliography{sample-base}

\begin{thebibliography}{10}
\providecommand{\url}[1]{#1}
\csname url@samestyle\endcsname
\providecommand{\newblock}{\relax}
\providecommand{\bibinfo}[2]{#2}
\providecommand{\BIBentrySTDinterwordspacing}{\spaceskip=0pt\relax}
\providecommand{\BIBentryALTinterwordstretchfactor}{4}
\providecommand{\BIBentryALTinterwordspacing}{\spaceskip=\fontdimen2\font plus
\BIBentryALTinterwordstretchfactor\fontdimen3\font minus \fontdimen4\font\relax}
\providecommand{\BIBforeignlanguage}[2]{{%
\expandafter\ifx\csname l@#1\endcsname\relax
\typeout{** WARNING: IEEEtran.bst: No hyphenation pattern has been}%
\typeout{** loaded for the language `#1'. Using the pattern for}%
\typeout{** the default language instead.}%
\else
\language=\csname l@#1\endcsname
\fi
#2}}
\providecommand{\BIBdecl}{\relax}
\BIBdecl

\bibitem{ying2018graph}
R.~Ying, R.~He, K.~Chen, P.~Eksombatchai, W.~L. Hamilton, and J.~Leskovec, ``Graph convolutional neural networks for web-scale recommender systems,'' in \emph{Proceedings of the 24th ACM SIGKDD international conference on knowledge discovery \& data mining}, 2018, pp. 974--983.

\bibitem{he2020lightgcn}
X.~He, K.~Deng, X.~Wang, Y.~Li, Y.~Zhang, and M.~Wang, ``Lightgcn: Simplifying and powering graph convolution network for recommendation,'' in \emph{Proceedings of the 43rd International ACM SIGIR conference on research and development in Information Retrieval}, 2020, pp. 639--648.

\bibitem{achiam2023gpt}
J.~Achiam, S.~Adler, S.~Agarwal, L.~Ahmad, I.~Akkaya, F.~L. Aleman, D.~Almeida, J.~Altenschmidt, S.~Altman, S.~Anadkat \emph{et~al.}, ``Gpt-4 technical report,'' \emph{arXiv preprint arXiv:2303.08774}, 2023.

\bibitem{wang2024visionllm}
W.~Wang, Z.~Chen, X.~Chen, J.~Wu, X.~Zhu, G.~Zeng, P.~Luo, T.~Lu, J.~Zhou, Y.~Qiao \emph{et~al.}, ``Visionllm: Large language model is also an open-ended decoder for vision-centric tasks,'' \emph{Advances in Neural Information Processing Systems}, vol.~36, 2024.

\bibitem{li2023blip}
J.~Li, D.~Li, S.~Savarese, and S.~Hoi, ``Blip-2: Bootstrapping language-image pre-training with frozen image encoders and large language models,'' in \emph{International conference on machine learning}.\hskip 1em plus 0.5em minus 0.4em\relax PMLR, 2023, pp. 19\,730--19\,742.

\bibitem{thirunavukarasu2023large}
A.~J. Thirunavukarasu, D.~S.~J. Ting, K.~Elangovan, L.~Gutierrez, T.~F. Tan, and D.~S.~W. Ting, ``Large language models in medicine,'' \emph{Nature medicine}, vol.~29, no.~8, pp. 1930--1940, 2023.

\bibitem{guo2023automated}
L.~Guo, C.~Wang, X.~Wang, L.~Zhu, and H.~Yin, ``Automated prompting for non-overlapping cross-domain sequential recommendation,'' \emph{arXiv preprint arXiv:2304.04218}, 2023.

\bibitem{wu2024exploring}
L.~Wu, Z.~Qiu, Z.~Zheng, H.~Zhu, and E.~Chen, ``Exploring large language model for graph data understanding in online job recommendations,'' in \emph{Proceedings of the AAAI Conference on Artificial Intelligence}, vol.~38, no.~8, 2024, pp. 9178--9186.

\bibitem{li2023personalized}
L.~Li, Y.~Zhang, and L.~Chen, ``Personalized prompt learning for explainable recommendation,'' \emph{ACM Transactions on Information Systems}, vol.~41, no.~4, pp. 1--26, 2023.

\bibitem{zhang2023collm}
Y.~Zhang, F.~Feng, J.~Zhang, K.~Bao, Q.~Wang, and X.~He, ``Collm: Integrating collaborative embeddings into large language models for recommendation,'' \emph{arXiv preprint arXiv:2310.19488}, 2023.

\bibitem{liao2024llara}
J.~Liao, S.~Li, Z.~Yang, J.~Wu, Y.~Yuan, X.~Wang, and X.~He, ``Llara: Large language-recommendation assistant,'' in \emph{Proceedings of the 47th International ACM SIGIR Conference on Research and Development in Information Retrieval}, 2024, pp. 1785--1795.

\bibitem{lin2024data}
X.~Lin, W.~Wang, Y.~Li, S.~Yang, F.~Feng, Y.~Wei, and T.-S. Chua, ``Data-efficient fine-tuning for llm-based recommendation,'' in \emph{Proceedings of the 47th International ACM SIGIR Conference on Research and Development in Information Retrieval}, 2024, pp. 365--374.

\bibitem{wu2024personalized}
Y.~Wu, R.~Xie, Y.~Zhu, F.~Zhuang, X.~Zhang, L.~Lin, and Q.~He, ``Personalized prompt for sequential recommendation,'' \emph{IEEE Transactions on Knowledge and Data Engineering}, 2024.

\bibitem{ding2023parameter}
N.~Ding, Y.~Qin, G.~Yang, F.~Wei, Z.~Yang, Y.~Su, S.~Hu, Y.~Chen, C.-M. Chan, W.~Chen \emph{et~al.}, ``Parameter-efficient fine-tuning of large-scale pre-trained language models,'' \emph{Nature Machine Intelligence}, vol.~5, no.~3, pp. 220--235, 2023.

\bibitem{lialin2023scaling}
V.~Lialin, V.~Deshpande, and A.~Rumshisky, ``Scaling down to scale up: A guide to parameter-efficient fine-tuning,'' \emph{arXiv preprint arXiv:2303.15647}, 2023.

\bibitem{li2023e4srec}
X.~Li, C.~Chen, X.~Zhao, Y.~Zhang, and C.~Xing, ``E4srec: An elegant effective efficient extensible solution of large language models for sequential recommendation,'' \emph{arXiv preprint arXiv:2312.02443}, 2023.

\bibitem{hidasi2015session}
B.~Hidasi, ``Session-based recommendations with recurrent neural networks,'' \emph{arXiv preprint arXiv:1511.06939}, 2015.

\bibitem{kang2018self}
W.-C. Kang and J.~McAuley, ``Self-attentive sequential recommendation,'' in \emph{2018 IEEE international conference on data mining (ICDM)}.\hskip 1em plus 0.5em minus 0.4em\relax IEEE, 2018, pp. 197--206.

\bibitem{wang2019neural}
X.~Wang, X.~He, M.~Wang, F.~Feng, and T.-S. Chua, ``Neural graph collaborative filtering,'' in \emph{Proceedings of the 42nd international ACM SIGIR conference on Research and development in Information Retrieval}, 2019, pp. 165--174.

\bibitem{wu2022graph}
S.~Wu, F.~Sun, W.~Zhang, X.~Xie, and B.~Cui, ``Graph neural networks in recommender systems: a survey,'' \emph{ACM Computing Surveys}, vol.~55, no.~5, pp. 1--37, 2022.

\bibitem{zhou2020s3}
K.~Zhou, H.~Wang, W.~X. Zhao, Y.~Zhu, S.~Wang, F.~Zhang, Z.~Wang, and J.-R. Wen, ``S3-rec: Self-supervised learning for sequential recommendation with mutual information maximization,'' in \emph{Proceedings of the 29th ACM international conference on information \& knowledge management}, 2020, pp. 1893--1902.

\bibitem{lester2021power}
B.~Lester, R.~Al-Rfou, and N.~Constant, ``The power of scale for parameter-efficient prompt tuning,'' \emph{arXiv preprint arXiv:2104.08691}, 2021.

\bibitem{hu2021lora}
E.~J. Hu, Y.~Shen, P.~Wallis, Z.~Allen-Zhu, Y.~Li, S.~Wang, L.~Wang, and W.~Chen, ``Lora: Low-rank adaptation of large language models,'' \emph{arXiv preprint arXiv:2106.09685}, 2021.

\bibitem{wu2019session}
S.~Wu, Y.~Tang, Y.~Zhu, L.~Wang, X.~Xie, and T.~Tan, ``Session-based recommendation with graph neural networks,'' in \emph{Proceedings of the AAAI conference on artificial intelligence}, vol.~33, no.~01, 2019, pp. 346--353.

\bibitem{xu2019graph}
C.~Xu, P.~Zhao, Y.~Liu, V.~S. Sheng, J.~Xu, F.~Zhuang, J.~Fang, and X.~Zhou, ``Graph contextualized self-attention network for session-based recommendation.'' in \emph{IJCAI}, vol.~19, 2019, pp. 3940--3946.

\bibitem{yu2020enhancing}
J.~Yu, H.~Yin, J.~Li, M.~Gao, Z.~Huang, and L.~Cui, ``Enhancing social recommendation with adversarial graph convolutional networks,'' \emph{IEEE Transactions on knowledge and data engineering}, vol.~34, no.~8, pp. 3727--3739, 2020.

\bibitem{fan2019graph}
W.~Fan, Y.~Ma, Q.~Li, Y.~He, E.~Zhao, J.~Tang, and D.~Yin, ``Graph neural networks for social recommendation,'' in \emph{The world wide web conference}, 2019, pp. 417--426.

\bibitem{wang2019knowledge}
H.~Wang, M.~Zhao, X.~Xie, W.~Li, and M.~Guo, ``Knowledge graph convolutional networks for recommender systems,'' in \emph{The world wide web conference}, 2019, pp. 3307--3313.

\bibitem{wang2019kgat}
X.~Wang, X.~He, Y.~Cao, M.~Liu, and T.-S. Chua, ``Kgat: Knowledge graph attention network for recommendation,'' in \emph{Proceedings of the 25th ACM SIGKDD international conference on knowledge discovery \& data mining}, 2019, pp. 950--958.

\bibitem{he2016vbpr}
R.~He and J.~McAuley, ``Vbpr: visual bayesian personalized ranking from implicit feedback,'' in \emph{Proceedings of the AAAI conference on artificial intelligence}, vol.~30, no.~1, 2016.

\bibitem{wei2019mmgcn}
Y.~Wei, X.~Wang, L.~Nie, X.~He, R.~Hong, and T.-S. Chua, ``Mmgcn: Multi-modal graph convolution network for personalized recommendation of micro-video,'' in \emph{Proceedings of the 27th ACM international conference on multimedia}, 2019, pp. 1437--1445.

\bibitem{wei2020graph}
Y.~Wei, X.~Wang, L.~Nie, X.~He, and T.-S. Chua, ``Graph-refined convolutional network for multimedia recommendation with implicit feedback,'' in \emph{Proceedings of the 28th ACM international conference on multimedia}, 2020, pp. 3541--3549.

\bibitem{zhang2021mining}
J.~Zhang, Y.~Zhu, Q.~Liu, S.~Wu, S.~Wang, and L.~Wang, ``Mining latent structures for multimedia recommendation,'' in \emph{Proceedings of the 29th ACM international conference on multimedia}, 2021, pp. 3872--3880.

\bibitem{wang2021dualgnn}
Q.~Wang, Y.~Wei, J.~Yin, J.~Wu, X.~Song, and L.~Nie, ``Dualgnn: Dual graph neural network for multimedia recommendation,'' \emph{IEEE Transactions on Multimedia}, vol.~25, pp. 1074--1084, 2021.

\bibitem{yi2021multi}
J.~Yi and Z.~Chen, ``Multi-modal variational graph auto-encoder for recommendation systems,'' \emph{IEEE Transactions on Multimedia}, vol.~24, pp. 1067--1079, 2021.

\bibitem{zhou2023tale}
X.~Zhou and Z.~Shen, ``A tale of two graphs: Freezing and denoising graph structures for multimodal recommendation,'' in \emph{Proceedings of the 31st ACM International Conference on Multimedia}, 2023, pp. 935--943.

\bibitem{chen2017attentive}
J.~Chen, H.~Zhang, X.~He, L.~Nie, W.~Liu, and T.-S. Chua, ``Attentive collaborative filtering: Multimedia recommendation with item-and component-level attention,'' in \emph{Proceedings of the 40th International ACM SIGIR conference on Research and Development in Information Retrieval}, 2017, pp. 335--344.

\bibitem{chen2019personalized}
X.~Chen, H.~Chen, H.~Xu, Y.~Zhang, Y.~Cao, Z.~Qin, and H.~Zha, ``Personalized fashion recommendation with visual explanations based on multimodal attention network: Towards visually explainable recommendation,'' in \emph{Proceedings of the 42nd International ACM SIGIR Conference on Research and Development in Information Retrieval}, 2019, pp. 765--774.

\bibitem{wei2023multi}
W.~Wei, C.~Huang, L.~Xia, and C.~Zhang, ``Multi-modal self-supervised learning for recommendation,'' in \emph{Proceedings of the ACM Web Conference 2023}, 2023, pp. 790--800.

\bibitem{zhou2023bootstrap}
X.~Zhou, H.~Zhou, Y.~Liu, Z.~Zeng, C.~Miao, P.~Wang, Y.~You, and F.~Jiang, ``Bootstrap latent representations for multi-modal recommendation,'' in \emph{Proceedings of the ACM Web Conference 2023}, 2023, pp. 845--854.

\bibitem{tao2022self}
Z.~Tao, X.~Liu, Y.~Xia, X.~Wang, L.~Yang, X.~Huang, and T.-S. Chua, ``Self-supervised learning for multimedia recommendation,'' \emph{IEEE Transactions on Multimedia}, 2022.

\bibitem{guo2025mmhcl}
X.~Guo, T.~Zhang, F.~Wang, X.~Wang, X.~Zhang, X.~Liu, and Z.~Cui, ``Mmhcl: Multi-modal hypergraph contrastive learning for recommendation,'' \emph{arXiv preprint arXiv:2504.16576}, 2025.

\bibitem{guo2025m}
X.~Guo, T.~Zhang, Y.~Xue, C.~Wang, F.~Wang, and Z.~Cui, ``M 3 rec: Selective state space models with mixture-of-modality experts for multi-modal sequential recommendation,'' in \emph{ICASSP 2025-2025 IEEE International Conference on Acoustics, Speech and Signal Processing (ICASSP)}.\hskip 1em plus 0.5em minus 0.4em\relax IEEE, 2025, pp. 1--5.

\bibitem{liu2024alignrec}
Y.~Liu, K.~Zhang, X.~Ren, Y.~Huang, J.~Jin, Y.~Qin, R.~Su, R.~Xu, Y.~Yu, and W.~Zhang, ``Alignrec: Aligning and training in multimodal recommendations,'' in \emph{Proceedings of the 33rd ACM International Conference on Information and Knowledge Management}, 2024, pp. 1503--1512.

\bibitem{yu2025mind}
P.~Yu, Z.~Tan, G.~Lu, and B.-K. Bao, ``Mind individual information! principal graph learning for multimedia recommendation,'' in \emph{Proceedings of the AAAI Conference on Artificial Intelligence}, vol.~39, no.~12, 2025, pp. 13\,096--13\,105.

\bibitem{houlsby2019parameter}
N.~Houlsby, A.~Giurgiu, S.~Jastrzebski, B.~Morrone, Q.~De~Laroussilhe, A.~Gesmundo, M.~Attariyan, and S.~Gelly, ``Parameter-efficient transfer learning for nlp,'' in \emph{International conference on machine learning}.\hskip 1em plus 0.5em minus 0.4em\relax PMLR, 2019, pp. 2790--2799.

\bibitem{li2021prefix}
X.~L. Li and P.~Liang, ``Prefix-tuning: Optimizing continuous prompts for generation,'' \emph{arXiv preprint arXiv:2101.00190}, 2021.

\bibitem{sung2022lst}
Y.-L. Sung, J.~Cho, and M.~Bansal, ``Lst: Ladder side-tuning for parameter and memory efficient transfer learning,'' \emph{Advances in Neural Information Processing Systems}, vol.~35, pp. 12\,991--13\,005, 2022.

\bibitem{zaken2021bitfit}
E.~B. Zaken, S.~Ravfogel, and Y.~Goldberg, ``Bitfit: Simple parameter-efficient fine-tuning for transformer-based masked language-models,'' \emph{arXiv preprint arXiv:2106.10199}, 2021.

\bibitem{guo2020parameter}
D.~Guo, A.~M. Rush, and Y.~Kim, ``Parameter-efficient transfer learning with diff pruning,'' \emph{arXiv preprint arXiv:2012.07463}, 2020.

\bibitem{sung2021training}
Y.-L. Sung, V.~Nair, and C.~A. Raffel, ``Training neural networks with fixed sparse masks,'' \emph{Advances in Neural Information Processing Systems}, vol.~34, pp. 24\,193--24\,205, 2021.

\bibitem{aghajanyan2020intrinsic}
A.~Aghajanyan, L.~Zettlemoyer, and S.~Gupta, ``Intrinsic dimensionality explains the effectiveness of language model fine-tuning,'' \emph{arXiv preprint arXiv:2012.13255}, 2020.

\bibitem{he2021towards}
J.~He, C.~Zhou, X.~Ma, T.~Berg-Kirkpatrick, and G.~Neubig, ``Towards a unified view of parameter-efficient transfer learning,'' \emph{arXiv preprint arXiv:2110.04366}, 2021.

\bibitem{mao2021unipelt}
Y.~Mao, L.~Mathias, R.~Hou, A.~Almahairi, H.~Ma, J.~Han, W.-t. Yih, and M.~Khabsa, ``Unipelt: A unified framework for parameter-efficient language model tuning,'' \emph{arXiv preprint arXiv:2110.07577}, 2021.

\bibitem{vaswani2017attention}
A.~Vaswani, N.~Shazeer, N.~Parmar, J.~Uszkoreit, L.~Jones, A.~N. Gomez, L.~Kaiser, and I.~Polosukhin, ``Attention is all you need,'' \emph{Advances in neural information processing systems}, vol.~30, 2017.

\bibitem{fan2023recommender}
W.~Fan, Z.~Zhao, J.~Li, Y.~Liu, X.~Mei, Y.~Wang, J.~Tang, and Q.~Li, ``Recommender systems in the era of large language models (llms),'' \emph{arXiv preprint arXiv:2307.02046}, 2023.

\bibitem{zhao2023survey}
W.~X. Zhao, K.~Zhou, J.~Li, T.~Tang, X.~Wang, Y.~Hou, Y.~Min, B.~Zhang, J.~Zhang, Z.~Dong \emph{et~al.}, ``A survey of large language models,'' \emph{arXiv preprint arXiv:2303.18223}, 2023.

\bibitem{hou2024large}
Y.~Hou, J.~Zhang, Z.~Lin, H.~Lu, R.~Xie, J.~McAuley, and W.~X. Zhao, ``Large language models are zero-shot rankers for recommender systems,'' in \emph{European Conference on Information Retrieval}.\hskip 1em plus 0.5em minus 0.4em\relax Springer, 2024, pp. 364--381.

\bibitem{gao2023chat}
Y.~Gao, T.~Sheng, Y.~Xiang, Y.~Xiong, H.~Wang, and J.~Zhang, ``Chat-rec: Towards interactive and explainable llms-augmented recommender system,'' \emph{arXiv preprint arXiv:2303.14524}, 2023.

\bibitem{liu2023chatgpt}
J.~Liu, C.~Liu, R.~Lv, K.~Zhou, and Y.~Zhang, ``Is chatgpt a good recommender? a preliminary study,'' \emph{arXiv preprint arXiv:2304.10149}, 2023.

\bibitem{zhang2023chatgpt}
J.~Zhang, K.~Bao, Y.~Zhang, W.~Wang, F.~Feng, and X.~He, ``Is chatgpt fair for recommendation? evaluating fairness in large language model recommendation,'' in \emph{Proceedings of the 17th ACM Conference on Recommender Systems}, 2023, pp. 993--999.

\bibitem{dai2023uncovering}
S.~Dai, N.~Shao, H.~Zhao, W.~Yu, Z.~Si, C.~Xu, Z.~Sun, X.~Zhang, and J.~Xu, ``Uncovering chatgpt’s capabilities in recommender systems,'' in \emph{Proceedings of the 17th ACM Conference on Recommender Systems}, 2023, pp. 1126--1132.

\bibitem{wang2023zero}
L.~Wang and E.-P. Lim, ``Zero-shot next-item recommendation using large pretrained language models,'' \emph{arXiv preprint arXiv:2304.03153}, 2023.

\bibitem{ren2023representation}
X.~Ren, W.~Wei, L.~Xia, L.~Su, S.~Cheng, J.~Wang, D.~Yin, and C.~Huang, ``Representation learning with large language models for recommendation,'' \emph{arXiv preprint arXiv:2310.15950}, 2023.

\bibitem{he2023large}
Z.~He, Z.~Xie, R.~Jha, H.~Steck, D.~Liang, Y.~Feng, B.~P. Majumder, N.~Kallus, and J.~McAuley, ``Large language models as zero-shot conversational recommenders,'' in \emph{Proceedings of the 32nd ACM international conference on information and knowledge management}, 2023, pp. 720--730.

\bibitem{sanner2023large}
S.~Sanner, K.~Balog, F.~Radlinski, B.~Wedin, and L.~Dixon, ``Large language models are competitive near cold-start recommenders for language-and item-based preferences,'' in \emph{Proceedings of the 17th ACM conference on recommender systems}, 2023, pp. 890--896.

\bibitem{bao2023tallrec}
K.~Bao, J.~Zhang, Y.~Zhang, W.~Wang, F.~Feng, and X.~He, ``Tallrec: An effective and efficient tuning framework to align large language model with recommendation,'' in \emph{Proceedings of the 17th ACM Conference on Recommender Systems}, 2023, pp. 1007--1014.

\bibitem{li2023prompt}
L.~Li, Y.~Zhang, and L.~Chen, ``Prompt distillation for efficient llm-based recommendation,'' in \emph{Proceedings of the 32nd ACM International Conference on Information and Knowledge Management}, 2023, pp. 1348--1357.

\bibitem{cui2022m6}
Z.~Cui, J.~Ma, C.~Zhou, J.~Zhou, and H.~Yang, ``M6-rec: Generative pretrained language models are open-ended recommender systems,'' \emph{arXiv preprint arXiv:2205.08084}, 2022.

\bibitem{wang2022towards}
X.~Wang, K.~Zhou, J.-R. Wen, and W.~X. Zhao, ``Towards unified conversational recommender systems via knowledge-enhanced prompt learning,'' in \emph{Proceedings of the 28th ACM SIGKDD Conference on Knowledge Discovery and Data Mining}, 2022, pp. 1929--1937.

\bibitem{zhang2024notellm}
C.~Zhang, S.~Wu, H.~Zhang, T.~Xu, Y.~Gao, Y.~Hu, and E.~Chen, ``Notellm: A retrievable large language model for note recommendation,'' in \emph{Companion Proceedings of the ACM on Web Conference 2024}, 2024, pp. 170--179.

\bibitem{feng2019hypergraph}
Y.~Feng, H.~You, Z.~Zhang, R.~Ji, and Y.~Gao, ``Hypergraph neural networks,'' in \emph{Proceedings of the AAAI conference on artificial intelligence}, vol.~33, no.~01, 2019, pp. 3558--3565.

\bibitem{glorot2010understanding}
X.~Glorot and Y.~Bengio, ``Understanding the difficulty of training deep feedforward neural networks,'' in \emph{Proceedings of the thirteenth international conference on artificial intelligence and statistics}.\hskip 1em plus 0.5em minus 0.4em\relax JMLR Workshop and Conference Proceedings, 2010, pp. 249--256.

\bibitem{wang2020second}
Z.~Wang and S.~Ji, ``Second-order pooling for graph neural networks,'' \emph{IEEE Transactions on Pattern Analysis and Machine Intelligence}, vol.~45, no.~6, pp. 6870--6880, 2020.

\bibitem{rendle2012bpr}
S.~Rendle, C.~Freudenthaler, Z.~Gantner, and L.~Schmidt-Thieme, ``Bpr: Bayesian personalized ranking from implicit feedback,'' \emph{arXiv preprint arXiv:1205.2618}, 2012.

\bibitem{mcauley2015image}
J.~McAuley, C.~Targett, Q.~Shi, and A.~Van Den~Hengel, ``Image-based recommendations on styles and substitutes,'' in \emph{Proceedings of the 38th international ACM SIGIR conference on research and development in information retrieval}, 2015, pp. 43--52.

\bibitem{DBLP:journals/corr/abs-1810-04805}
\BIBentryALTinterwordspacing
J.~Devlin, M.~Chang, K.~Lee, and K.~Toutanova, ``{BERT:} pre-training of deep bidirectional transformers for language understanding,'' \emph{CoRR}, vol. abs/1810.04805, 2018. [Online]. Available: \url{http://arxiv.org/abs/1810.04805}
\BIBentrySTDinterwordspacing

\bibitem{radford2021learning}
A.~Radford, J.~W. Kim, C.~Hallacy, A.~Ramesh, G.~Goh, S.~Agarwal, G.~Sastry, A.~Askell, P.~Mishkin, J.~Clark \emph{et~al.}, ``Learning transferable visual models from natural language supervision,'' in \emph{International conference on machine learning}.\hskip 1em plus 0.5em minus 0.4em\relax PMLR, 2021, pp. 8748--8763.

\bibitem{guo2024lgmrec}
Z.~Guo, J.~Li, G.~Li, C.~Wang, S.~Shi, and B.~Ruan, ``Lgmrec: Local and global graph learning for multimodal recommendation,'' in \emph{Proceedings of the AAAI Conference on Artificial Intelligence}, vol.~38, no.~8, 2024, pp. 8454--8462.

\bibitem{touvron2023llama}
H.~Touvron, L.~Martin, K.~Stone, P.~Albert, A.~Almahairi, Y.~Babaei, N.~Bashlykov, S.~Batra, P.~Bhargava, S.~Bhosale \emph{et~al.}, ``Llama 2: Open foundation and fine-tuned chat models,'' \emph{arXiv preprint arXiv:2307.09288}, 2023.

\bibitem{kingma2014adam}
D.~P. Kingma and J.~Ba, ``Adam: A method for stochastic optimization,'' \emph{arXiv preprint arXiv:1412.6980}, 2014.

\end{thebibliography}

\end{document}